\documentclass[prd,preprint,superscriptaddress,preprintnumbers,eqsecnum,nofootinbib,nobibnotes]{revtex4}
\usepackage{framed}
\usepackage{bbold}
\usepackage{slashed}
\usepackage{graphicx}
\usepackage{amsmath}
\usepackage{amssymb}
\usepackage{epsfig}
\usepackage{hhline}
\usepackage{soul}
\usepackage{tabularx,pifont,multirow}
\usepackage{enumitem}
\usepackage{colordvi}
\usepackage{soul}
\usepackage{xcolor}
\usepackage{yfonts}

\newcommand{\be}{\begin{equation}}
\newcommand{\ee}{\end{equation}}
\newcommand{\bea}{\begin{eqnarray}}
\newcommand{\eea}{\end{eqnarray}}

\definecolor{darkgreen}{rgb}{0,0.3,0.05}
\makeatletter                                                         %
\newcommand*\rel@kern[1]{\kern#1\dimexpr\macc@kerna}                  %
\newcommand*\widebar[1]{                                              %
  \begingroup                                                         %
  \def\mathaccent##1##2{                                              %
    \rel@kern{0.8}                                                    %
    \overline{\rel@kern{-0.8}\macc@nucleus\rel@kern{0.2}}             %
    \rel@kern{-0.2}                                                   %
  }                                                                   %
  \macc@depth\@ne                                                     %
  \let\math@bgroup\@empty \let\math@egroup\macc@set@skewchar          %
  \mathsurround\z@ \frozen@everymath{\mathgroup\macc@group\relax}     %
  \macc@set@skewchar\relax                                            %
  \let\mathaccentV\macc@nested@a                                      %
  \macc@nested@a\relax111{#1}                                         %
  \endgroup                                                           %
}                                                                     %
\makeatother                                                          %

\begin{document}

\leftline{KCL-PH-TH/2017-{\bf 61}, \,IFIC/17-58}

%

\title{Singular lensing from the scattering on special space-time defects}


\bigskip

\author{Nick E.~Mavromatos}
\affiliation{Department of Theoretical Physics 
  and IFIC, University of Valencia - CSIC, E-46100, Spain}
\affiliation{  
Theoretical Particle Physics and Cosmology Group, Department of Physics, King's College London,  
Strand, London WC2R 2LS, UK.}

\author{Joannis Papavassiliou}
\affiliation{Department of Theoretical Physics 
  and IFIC, University of Valencia - CSIC, E-46100, Spain}


\begin{abstract}
  
It  is well  known that  certain special  classes of  self-gravitating
point-like defects, such  as global (non gauged)  monopoles, give rise
to non-asymptotically  flat space-times  characterized by  solid angle
deficits,  whose  size  depends  on  the  details  of  the  underlying
microscopic models.  The scattering of electrically  neutral particles
on such space-times  is described by amplitudes  that exhibit resonant
behaviour when the  scattering and deficit angles  coincide.  This, in
turn,  leads to  ring-like  structures where  the  cross sections  are
formally divergent  (``singular lensing'').  In this  work, we revisit
this particular phenomenon, with the twofold purpose of  placing it  in a
contemporary and more general context,  in view of renewed interest in
the theory and general phenomenology  of such defects, and, 
more importantly,  of addressing  certain subtleties that appear in
the particular  computation that  leads to the  aforementioned effect.
In particular, by adopting a specific regularization procedure for the
formally infinite Legendre series encountered, we manage to ensure the
recovery of  the Minkowski space-time,  and thus the  disappearance of
the lensing  phenomenon, in the  no-defect limit, and the  validity of
the optical theorem for the elastic total cross section.
In addition, the singular nature of the phenomenon  
is confirmed by means of an alternative calculation, which, unlike the
original approach, makes no use
of the generating function of the Legendre polynomials, but rather exploits 
the asymptotic properties of the Fresnel integrals.

\end{abstract}
\maketitle

\section{Introduction\label{sec:Intro}}

The presence of space-time defects in a physical system always presents interesting but also challenging aspects from both physical and technical points of view.
By ``defects'' we mean field theoretic entities, either point-like or with (solitonic) structure, which exhibit singularities at a given point in space-time.
Scattering of other particles on such backgrounds leads to interesting and non-trivial effects, both at the classical and the quantum level.
Magnetic Dirac monopoles~\cite{dirac} are a prototypical example of such defects~\cite{shir}, namely point-like objects
characterized by singular electromagnetic potentials at their origin, which induce singular (gauge invariant) magnetic fields.
The intensities of these latter fields are proportional to the magnetic charge, which, due to Dirac's quantization condition, 
is a half-integer multiple of the inverse of the electric charge. 
For curved space times, black holes are singularities of the gravitational field, leading to singularities in curvature invariants, which constitute, in some sense, the analogue of the infinite gauge invariant magnetic field intensity of the magnetic monopole case.
The embedding of monopoles in curved space-times results in interesting and highly non trivial circumstances,
\emph{e.g.}, black hole horizons enveloping the monopole~\cite{gravmon}.

Scattering of particles off magnetic monopoles and/or black holes is well studied by now. It is worth mentioning that, as far as magnetic monopoles are concerned, both classical and quantum scattering have revealed interesting features on the motion of a particle, which dates  back to the work of Poincar\`e~\cite{poincare}.
In an attempt to understand the focusing motion of electrons in a cathodic tube in the Birkeland experiment~\cite{birkeland} in the presence of an external electromagnet,
Poincar\`e used the notion of a magnetic ``monopole'', by interpreting the electromagnet as the source
of a singular magnetic field (isolated ``north magnetic pole'').  Poincar\'e discovered that the classical trajectory of an electron moving towards the magnetic pole follows the geodesics on a \emph{cone}, whose appex is located in the position of the isolated magnetic pole, and whose  generatrix is the axis of the angular momentum $\vec J $.
The angle of the cone 
is given by ${\rm cot}\theta = e g /|\vec J|$, where $g$ is the magnetic charge, and $e$ the charge of the electron. 
If a ring of such electrons is considered, then Poincar\'e's work demonstrated that their trajectories will focus towards the monopole, up to a minimum distance, before scattered away, thereby providing an ``explanation'' of the results of the experiment of Birkeland~\cite{birkeland}, who had also conjectured that the electrons in his experiment somehow were following the magnetic field lines. 
Dirac introduced the concept and our modern understanding of a magnetic monopole explicitly some thirty years later~\cite{dirac}.  Subsequently, `t Hooft and Polyakov~\cite{hpmono} put the monopole in 
the context of spontaneously broken non-abelian gauge theories, but this monopole has (solitonic) structure, in contrast to the point-like Dirac one.
The quantum scattering of particles off such magnetic monopoles were discussed in a plethora of works so far~\cite{monscat}.

Another kind of defect is the one proposed in \cite{vilenkin}, arising in spontaneously broken SO(3) internal isospin \emph{global} symmetry, which, in contrast to the ordinary monopoles (\emph{e.g.}, `t Hooft-Polyakov~\cite{hpmono}), when embedded in a curved space-time induces a conical singularity, in the sense of an \emph{angular deficit} proportional to the relevant vacuum expectation value responsible for the breaking of the symmetry. 
A string-inspired extension of the model of \cite{vilenkin}, in which the global monopole can induce a magnetic monopole,  
has also been discussed in~\cite{sarben}.  
In addition, space-times with angular deficits appear in models of three-spatial-dimensional Dirichlet brane Universes, moving in higher-dimensional bulk spaces. The latter contain populations of quantum fluctuating point-like D0-brane defects (D-foam), which can be bounded on the brane worlds, thus providing a ``medium'' in which quantum matter propagates~\cite{dfoam}.  The recoil fluctuations of such defects result in asymptotic space times with angular surpluses~\cite{recoil}. 

In ref.~\cite{mazur}, the quantum scattering of neutral scalar massless particles off global monopoles~\cite{vilenkin} has been considered. Given that the main interest of that work was the asymptotic features of the elastic collision, far away from the position of the defect, the study was restricted only in flat space-times but with the angular deficit induced by the defect; the latter was the only trace of the underlying complicated microscopic dynamics. The characteristic effect found was a ring-like angular region (in the forward direction) where the scattering amplitude and, hence, the elastic cross section, become very large (formally divergent). In what follows we shall refer to this phenomenon as ``lensing''. 

The analysis of \cite{mazur} is fairly generic and does not depend on the particular kind of defect that causes the conical singularity of space-time; in fact, the results are expected to hold also for the other kind of defects we mentioned above, namely the D-particle foam, which may have interesting implications in dark matter searches, in view of the r\^ole of the D-particles as dark matter candidates~\cite{sakell}. 
Generalizations of the results of \cite{mazur} to fermions 
have been presented in \cite{others1},
and charged massive particles (with the charge appearing only in self-interaction potential) in \cite{others2}. The comparison with the case of scattering off cosmic strings~\cite{cosmic} has been given in \cite{lousto}.
In this work we revisit the scalar massless case of \cite{mazur}. Our purpose is twofold. First, to put it in a contemporary and more general context, in view of renewed interest in the general phenomenology of such defects, ranging from cosmological and astrophysical observations~\cite{hindmarsh} to specific (magnetic monopole) searches in current collider experiments~\cite{atlasmon,moedal}. Second, and most important, to address  certain subtle and physically crucial issues, which appeared  in the particular computation that leads to the aforementioned effect. Specifically, our current study reveals that the aforementioned effect of \cite{mazur} was not an artefact of the formal manipulations employed, but 
persists our more rigorous treatment involving proper regularization of the pertinent Legendre polynomial series.  Moreover, this novel procedure guarantees leads a smooth recovery of the vanishing of the effect in the no-defect limit (\emph{i.e.} flat Minkowski space-time). 
This was one of the important missing ingredients in all previous analyses of the subject, where such a limit could not be recovered. In addition, the validity of the optical theorem, which is a direct consequence of unitarity (assumed to hold in the asymptotic region far away from the defect), is established through a non-trivial regularization procedure, whereby cut-off dependent quantities are judiciously employed~\footnote{The validity of the optical theorem in this context has been questioned in \cite{pusc}, on the ground of the curved nature of the space time in the presence of global defects.}. Last but not least, we present an alternative mathematical procedure 
for the evaluation of the scattering amplitude,  
which does not rely on the use of the generating function of the  Legendre polynomials, but makes instead extensive use of the
asymptotic properties of the Fresnel integrals. This particular construction confirms, in an independent and technically distinct way,  
the singular nature of the lensing phenomenon. 

The structure of the article is as follows. In section \ref{sec:models} we review certain representative
microscopic models that give rise to space-times with angular defects (deficit or surplus).
In the following section \ref{sec:lensing} we compute in detail the scattering amplitude of scalar massless particles on such defects and demonstrate the phenomenon of lensing. 
In section \ref{sec:nodeflimit} we discuss the transition to the no-deficit limit, by employing properly regularized Legenrde polynomial series, and an appropriate representation of the Dirac $\delta$-function distribution at the origin. In section  \ref{sec:pheno} we derive the lensing phenomenon at the level of the differential cross section.
This is followed, in section \ref{sec:appendix}, by a demonstration of the validity of the optical theorem within our regularization approach. Finally, in section \ref{sec:concl} we present our conclusions and discuss potential phenomenological applications, both cosmic and at particle colliders.

\section{Microscopic systems inducing space-time defects \label{sec:models}}
In this section we discuss certain microscopic models that may give
rise to space-time defects. 

\subsection{Global monopoles}
In \cite{vilenkin}, the  case of a self-gravitating global monopole  has been considered. The system consists of a triplet of Higgs-like scalar fields, $\chi^a$, $a=1,2,3$, which spontaneously break SO(3) symmetry to a global $U(1)$, through a vacuum expectation value (v.e.v.) $\eta$; however, 
the scalar fields do not couple to a gauge field, hence the difference from the standard `t Hooft-Polyakov monopoles~\cite{hpmono}.
The Lagrangian of the system, when placed in a curved space time with metric tensor $g_{\mu\nu}$ and Ricci scalar curvature $R$ reads
\begin{equation}
L=\left(-g\right)^{1/2}\left\{ \frac{1}{2}\partial_{\mu}\chi^{a}\partial^{\mu}\chi^{a}-\frac{\lambda}{4}\left(\chi^{a}\chi^{a}-\eta^{2}\right)^{2} -R\right\} \label{eq:gm}\,,
\end{equation}
where $g=\det\left(g_{\mu\nu}\right)$ is the metric determinant, and $\lambda > 0$ is the Higgs-like-field self-interaction coupling.

As a consequence of Goldstone's theorem, such monopoles have massless Goldstone modes associated with them, which have energy densities that scale like $1/r^2$ with the radial distance from the monopole core. This results in a linear divergence of the monopole total energy (mass), which is a characteristic feature of such solutions, in a way similar to the linearly divergent energy of a cosmic string.
In the original work of \cite{vilenkin} only estimates of the total monopole mass have been given, by considering the solution in the exterior of the monopole core, whose size in flat space time has been estimated to be of order 
 \begin{equation}\label{coresize}
 \delta \sim \lambda^{-1/2} \, \eta^{-1}~, 
 \end{equation}
 leading to a heuristic mass estimate of order 
\begin{equation}\label{heurmass}
M_{\rm core} \sim  \delta^3 \, \lambda \, \eta^4 = \lambda^{-1} \eta ~.
\end{equation}  
The presence of the monopole curves the space-time exterior, and these estimates, even the concept of the mass of the global monopole,  have to be rethought.
The main argument of \cite{vilenkin} was that gravitational effects are weak for $\eta \ll M_{\rm P}$, the Planck mass; this is certainly the case when $\eta$ is of order of a few TeV, the scale of relevance for new physics searches at LHC (however it should be noted that the scalar triplet  field $\chi^a, a=1,\dots 3$ does not represent the Higgs field of the Standard Model. It signifies new physics, an issue we shall come back to it later on in the article). 
In this sense, the authors of \cite{vilenkin} argued that the flat space-time estimates for the core mass might still be valid, as an order of magnitude estimate. Outside the monopole core, they used approximate asymptotic analysis of the Einstein equations, 
\begin{equation}\label{einsteq}
R_{\mu\nu} - \frac{1}{2} g_{\mu\nu} R = 8\pi G_{N} \, T^\chi_{\mu\nu}\,,
\end{equation}
where $T_{\mu\nu}^\chi$ is the matter stress tensor derived from the Lagrangian (\ref{eq:gm}) and the equations of motion for the scalar fields $\chi^a$, $a=1,2,3$ . The scalar field configuration for a global monopole is~\cite{vilenkin}
\begin{equation}\label{eq:gmf}
\chi^a = \eta \, f(r) \, \frac{x^a}{r}~, a=1,2,3 
\end{equation}
where $x^a$ are spatial Cartesian coordinates, $r = \sqrt{x^a x^a }$ is the radial distance, and $f(r) \to 1 $ for  $r \gg \delta $. So 
at such large distances, the amplitude squared of the scalar field triplet approaches the square of the vacuum expectation value $\eta$, $\chi^a \chi^a \to \eta^2$.
In fact, the reader may recognize the similarity between the expression (\ref{eq:gmf}) and  the corresponding one for the `t Hooft-Polyakov monopole,
although, as we explained above, the underlying physics between the two problems is entirely different. 

As was argued in \cite{vilenkin}, due to the symmetry breaking and the linearly divergent energy of the global monopole, 
the space-time {\it differs} from the standard, asymptotically flat Schwarzschild metric corresponding to a massive object with mass $M_{\rm core}$ (assuming that all the mass of the monopole is concentrated in the core's interior) when $r \gg \delta $; specifically, 
\begin{equation}\label{asymptmetric}
ds^2  = -\left( 1 - 8\pi \, G_{\rm N} \eta^2 - \frac{2 G_{N}\, M_{\rm core}}{r} \right) dt^2 - \frac{dr^2}{1 + 8\pi \, G_{N} \eta^2 - \frac{2 G_{\rm N}\, M_{\rm core}}{r}} + r^2 \Big( d\theta^2 + {\rm sin}^2 \theta \, d\phi ^2 \Big)~, 
\end{equation}
where the signature $(-,+,+,+)$ was adopted for the metric, and 
$(r,\theta,\phi)$ denote the spherical coordinates.


In the asymptotic limit $ r \to \infty$, upon appropriate rescaling of the time
\mbox{$t \to (1 - 8\pi \, G_{N} \eta^2)^{-1/2} \, t^\prime $}, and radial coordinate $r$, $r \to (1 - 8\pi \, G_{N} \eta^2)^{1/2}\, r^\prime$, the space-time (\ref{asympt}) becomes 
\begin{equation}\label{defspace}
ds^2  = -d{t^\prime}^2 + d{r^\prime}^2 +   \left( 1 - 8\pi \, G_{N} \eta^2 \right)\, {r^\prime}^2 \left( d\theta^2 + {\rm sin}^2 \theta \, d\phi ^2 \right)~, \quad r \gg \delta~,
\end{equation}
that is, it would formally resemble a Minkowski space-time but with a conical \emph{deficit solid angle} 
\be\label{deficit}
\Delta \Omega = 8\pi G_{N} \, \eta^2~. 
\ee
The existence of the deficit (\ref{deficit}) implies that the space-time (\ref{defspace}) (or, equivalently, (\ref{asympt}))  is not flat, since the scalar curvature behaves, on account of (\ref{einsteq}) and (\ref{eq:gm}),  as 
\begin{equation}
R \propto \frac{16\pi \, G_N \, \eta^2}{r^2}~.
\end{equation} 
The reader should note that in the unbroken phase $\eta = 0$, where the defect is massless, in view of (\ref{heurmass}), the space-time (\ref{defspace}) (or (\ref{asympt})) becomes the ordinary flat Minkowski space-time.\footnote{In the current work we do not comment on the stability of the global monopole configuration. A debate on this issue is still ongoing~\cite{debate}.} 

The presence of a monopole-induced deficit solid angle can have important physical consequences for scattering processes in such space-times. Indeed, 
as shown first in \cite{mazur}, for scalar neutral particles, and was generalised to fermions in~\cite{others1} and charged particles in~\cite{others2}, the quantum mechanical amplitude
describing the scattering of the particle off the defect in the space-time (\ref{defspace}) is very large for
regions of the (forward) scattering angle of order of the deficit angle (or equivalently the squared ratio of the monopole mass to the Planck mass). In this sense the defect acts as a focusing object for the scattering of particles off it. 

We mention at this stage that a discussion of  the phenomenon for scalar particles  was also presented in \cite{lousto}, independently of the earlier work of \cite{mazur}. In that work, the additional feature of moving defects (at ultra-galactic speeds) has been considered. Moreover, a regularization of some of the singular results of \cite{mazur}, in the limit where the defect is absent, has been attempted in \cite{lousto}, but as pointed out in \cite{pusc}, there were some algebraic errors which rendered some of those results inconsistent. 
 It is the purpose of the current work to address carefully such issues, before discussing the phenomenology of the effect in a modern context, where defects that can induce the asymptotically non-flat space times  (\ref{defspace}) are in principle producible at current colliders, such as the Large Hadron Collider (LHC, CERN), within the framework of new physics models, provided their masses of are of the order of a few TeV. 

 It should be stressed that the above property of the defect acting as a lens of scattered particles is independent of the details of the self-gravitating monopole solutions, and is due only to the existence of the deficit angle in the space-time (\ref{defspace}). In this respect we mention that, subsequent to the work of \cite{vilenkin}, more detailed analysis of the gravitational back reaction effects of such defects has been performed in \cite{negative}, by requiring a matching of the solutions of the non-linear coupled system of gravitational and matter equations at the core radius.\footnote{The motivation for using the above matching comes from the observation that, at the origin ($r \to 0$), the Higgs potential for the scalars leads to a cosmological constant $\propto \eta^4$, since any ``matter'' scalar  fields go to zero. However, if a black hole or other geometric singularity is present as $r \to 0$, the space-time is different for small $r$ ($r \to 0$), and 
the arguments leading to negative mass would then not hold.}

 In this way the latter can be 
determined dynamically, rather than heuristically from flat space-time arguments as in \cite{vilenkin}. 
Indeed, in \cite{negative}, the core radius $r_c = 2 \, \lambda^{-1/2}\, \eta^{-1}$ for the self-gravitating solution was found by matching an exterior  Schwarzschild-like metric 
\be
\label{negspace}
ds^2 = -\left(1 - 8\pi G_{N}\, \eta^2 \, - \frac{2\, G_N\, M}{r} \right) dt^2 +\left(1 - 8\pi G_{N}\, \eta^2 \,
+ \frac{2\, G_N\, M}{r} \right)^{-1}dr^2 - r^2 \, d\Omega^2 \,, 
 \ee
 to an interior local de Sitter metric 
\be\label{interiornegspace}
ds^2 = -\left(1 - \mathcal{H}^2 \, r^2 \right) dt^2 + \left(1 - \mathcal{H}^2 \, r^2 \right)^{-1} dr^2 + r^2 \, d\Omega^2 \,, 
\ee
where $M$ denotes the monopole mass and $\mathcal{H}^2 = \frac{8\pi G_{N}\, \lambda \, \eta^4}{12}$ the de Sitter parameter~\footnote{In fact, in \cite{bronnikov}, the authors presented a classification of space-times arising from self-gravitating global monopoles in field theories with only a triplet of Higgs-type scalar fields and Ricci scalar curvature. The conclusion is that, under the requirement of  \emph{regularity at the centre} of the monopole, as in \cite{negative}, and independently of the shape of the Higgs potential, the metric can contain at most one horizon, and, in case there is an horizon, the global space-time structure is that of a de Sitter space-time.}.

Unfortunately, such a matching yields a negative mass for the monopole, \mbox{$M\sim -6\pi \lambda^{-1/2} \eta <0$}.
The interpretation of this sign in~\cite{negative} is based on  the repulsive nature of gravity induced by the vacuum-energy ${\mathcal H}^2$ provided by the global monopole.
Moreover, it has been argued in \cite{negative} that this interpretation is consistent with the monopole being an entity with complicated structure rather than an elementary particle-like excitation. Even though a monopole 
with a negative mass is of no relevance to collider physics,
the scattering of particles in the resulting space-times (\ref{negspace}),(\ref{interiornegspace}) would 
still exhibit the lensing phenomenon of \cite{mazur}, as a result of the existence of the solid deficit angle (\ref{deficit})
in the asymptotic form of the metric (\ref{negspace}) far away from the monopole core. In view of the cosmological relevance
of the space-time (\ref{negspace}), (\ref{interiornegspace}),
the phenomenon suggested in \cite{mazur} may be useful in setting bounds for these defects in a cosmological context.

\subsection{Magnetic monopoles in models with antisymmetric tensor fields \label{sec:anttensor}} 

In \cite{sarben}, an extension of the global monopole model was preented, inspired from string theory, with dilaton $\Phi$ and antisymmetric tensor (spin 1) 
fields $B_{\mu\nu}=-B_{\nu\mu}$ present, which are known to characterize the massless gravitational multiplet of strings.
The model has also an electromagnetic field, $f_{\mu\nu}$, whose Maxwell tensor couples to the rest of the terms via appropriate dilaton terms 
\begin{eqnarray}
  L &=&\left(-g\right)^{1/2}\Big\{ \frac{1}{2}\partial_{\mu}\chi^{a}\partial^{\mu}\chi^{a}-\frac{\lambda}{4}\left(\chi^{a}\chi^{a}-\eta^{2}\right)^{2} -R
 \nonumber \\
 &+& \frac{1}{2}\partial_{\mu}\Phi\partial^{\mu}\Phi-V\left(\Phi\right) -\frac{1}{12}\, e^{-2\gamma \Phi}\, H_{\rho\mu\nu}H^{\varrho\mu\nu}-\frac{1}{4}\, e^{-\gamma \Phi}\, f_{\mu\nu}f^{\mu\nu}\Big\} \,,
\label{eq:13}
\end{eqnarray}
where $\gamma$ is a real constant, which in specific string theory models takes on the value $-1$, and the antisymmetric tensor field strength $H_{\rho\mu\nu}=\partial_{\left[\rho\right.}B_{\mu\left.\nu\right]}$, where the brackets $[\dots]$ denote total antisymmetrization of the respective indices.

As shown in \cite{sarben}, one may obtain monopole solutions with non-zero magnetic charge,
due to the coupling of $f_{\mu\nu}$ with the antisymmetric tensor field strength $H_{\mu\nu\rho}$, described by the dilaton equation of motion. 
In this case, the metric is that of Reissner-Nordstr\"om (RN) geometry due to the antisymmetric tensor and electromagnetic fields, with the r\^ole of the RN charge played by the magnetic charge of the monopole. The singular nature of the solution at $r \to 0$ invalidates the arguments of \cite{negative,bronnikov}, and one can obtain a positive mass for the magnetic monopole. The latter has been estimated in \cite{sarben} to be finite, for {\it strong coupling}, $\lambda \gg 1$, and 
assuming a kind of ``bag'' model for the monopole,
where the bulk of its mass comes from a thin shell of thickness $\alpha \, L$, $0 < \alpha \ll 1$ near the core radius $L$, 
\begin{eqnarray}\label{corer1}
\mathcal{M} &\sim & \int_{\rm shell ~thickness~(1-\alpha)L}\, \sqrt{-g}\, d^3x \, \left[ \frac{2\, W^2}{B\,r^2} + \frac{(b^\prime)^2}{4\, B A}     + \eta^2\, \left(\frac{f^2}{B  r^2} + \frac{(f^\prime)^2}{2B A} \right) + \frac{\lambda \, \eta^4}{4 B}\, (f^2 - 1)^2 \right] \nonumber \\
&\simeq &  \frac{1}{\alpha}\, (1-\alpha) \, \Big( 9\pi \zeta^2 + \frac{4\pi}{\lambda} \Big) \,  \frac{1}{L} + 4\pi \, \eta^2 \, (1 - \alpha ) \, L ~,
\end{eqnarray}
where the various functions depend only on the radial coordinate $r$. In the expression above, $A(r)$ and $B(r)$ are space-time metric functions, parametrizing components of the metric in the Schwarzschild system of coordinates $(t,r,\theta,\phi)$, with $ t $ the time and $r,\theta,\phi$ spherical coordinates, as follows:
$g_{00}=-B(r) , \, g_{rr} =A(r), \, g_{\theta\theta} =r^2, \, g_{\phi\phi}=r^2 {\rm sin}^2\, \theta $ in our signature convention; 
 $W(r) $ is a function associated with the solution for the Maxwell gauge field strength $f_{\mu\nu}$, such that its $\theta\phi$-component reads $f_{\theta\phi}=2r\,\,\theta\,{\rm sin}\theta \, W(r)$; $b(r) $ is a pseudoscalar field linked with the antisymmetric tensor field strength, which, in four space-time dimensions, can be expressed uniquely as $H_{\mu\nu\rho}=\epsilon_{\mu\nu\rho\sigma}\, \partial^\sigma b(x)$, and the ``prime'' denotes derivative with respect to $r$.  The monopole solution of \cite{sarben} is characterized by 
$b^\prime \left(r\right)=\frac{\zeta}{r^{2}}\sqrt{\frac{A\left(r\right)}{B\left(r\right)}}$, where $\sqrt{2}\,\zeta$
is its {\it magnetic charge}; finally, $f(r)$ characterizes the global-monopole scalar field Ansatz 
$\chi^{a}=\eta f\left(r\right)\frac{x^{a}}{r}~, \,  a=1,2,3 $, with $x^a$ Cartesian spatial coordinates, which are such that $\lim\limits_{r \to 0} f(r) = 0$ and $\lim\limits_{r \to \infty} f(r) = 1$. 

Minimization of (\ref{corer1}) with respect to $L=L_{\rm min}$ leads to a core size $L_{\rm min} \equiv L_c$ of order 
$L_c = 3 |\zeta |/2 \eta {\alpha}^{1/2}$, 
and thus to an estimate of the (positive) monopole mass,~\cite{sarben} 
\begin{eqnarray}\label{massfinal2}
\mathcal M \sim 12\pi\, \, {\alpha}^{-1/2} \, (1-\alpha) \, |\zeta |\, \eta  \, = \, (1-\alpha)\, 8\pi\, \eta^2 \, L_c \, > \, 0~, 
\end{eqnarray}
where $\alpha \ll 1$ is a number that must be determined from phenomenology.

The asymptotic $r \to \infty$ space-time induced by the self-gravitating global monopole, assumes the RN form~\cite{sarben}:
\be\label{RNform}
ds^2 = -\Big(1 - 8\pi G_{N}\, \eta^2 \, - \frac{2\, G_N\, M}{r}  + \frac{p_0}{r^2} \Big) dt^2 + \Big(1 - 8\pi G_{N}\, \eta^2 \, + \frac{2\, G_N\, M}{r} + \frac{p_0}{r^2}\Big)^{-1}
 dr^2 + r^2 \, d\Omega^2~,
 \ee
where $p_0 := 2\zeta^2 - 1/\lambda$.
 
The asymptotic space-time in (\ref{RNform}) has the angular deficit of the standard global solution (\ref{deficit}), but now the monopole is a highly ionising particle, on account of its magnetic charge. For sufficiently low v.e.v. $\eta $, such that the mass of the monopole (\ref{massfinal2}) is of order TeV, such objects can be produced at current colliders, but in monopole anti-monopole pairs. It should be remarked at this point that, if the monopole solutions have structure, such as the global-monopole-inspired solutions we are discussing in this work~\cite{vilenkin,negative,sarben}, or a `t Hooft-Polyakov monopole~\cite{hpmono}, then their production at colliders at zero or very low temperatures is expected to be extremely suppressed~\cite{drukier}.
However, abundant production of such objects may be expected~\cite{rajantie} in environments with high magnetic fields or high temperatures (such as neutron starts or in heavy ion collisions at colliders, such as the LHC), as a result of a Schwinger-like~\cite{Schwinger} thermal pair production mechanism from the vacuum, provided of course that the external conditions, \emph{e.g.} temperature, are such that one is in the broken phase of the SO(3) symmetry,  so that $\eta \ne 0$ (\emph{e.g.} the temperature is lower than the critical temperature for symmetry restoration).

\subsection{``Foam'' models: brane Universe with ensembles of D$0$-brane defects}
\begin{figure}[ht!]
\centering \includegraphics[width=7.5cm]{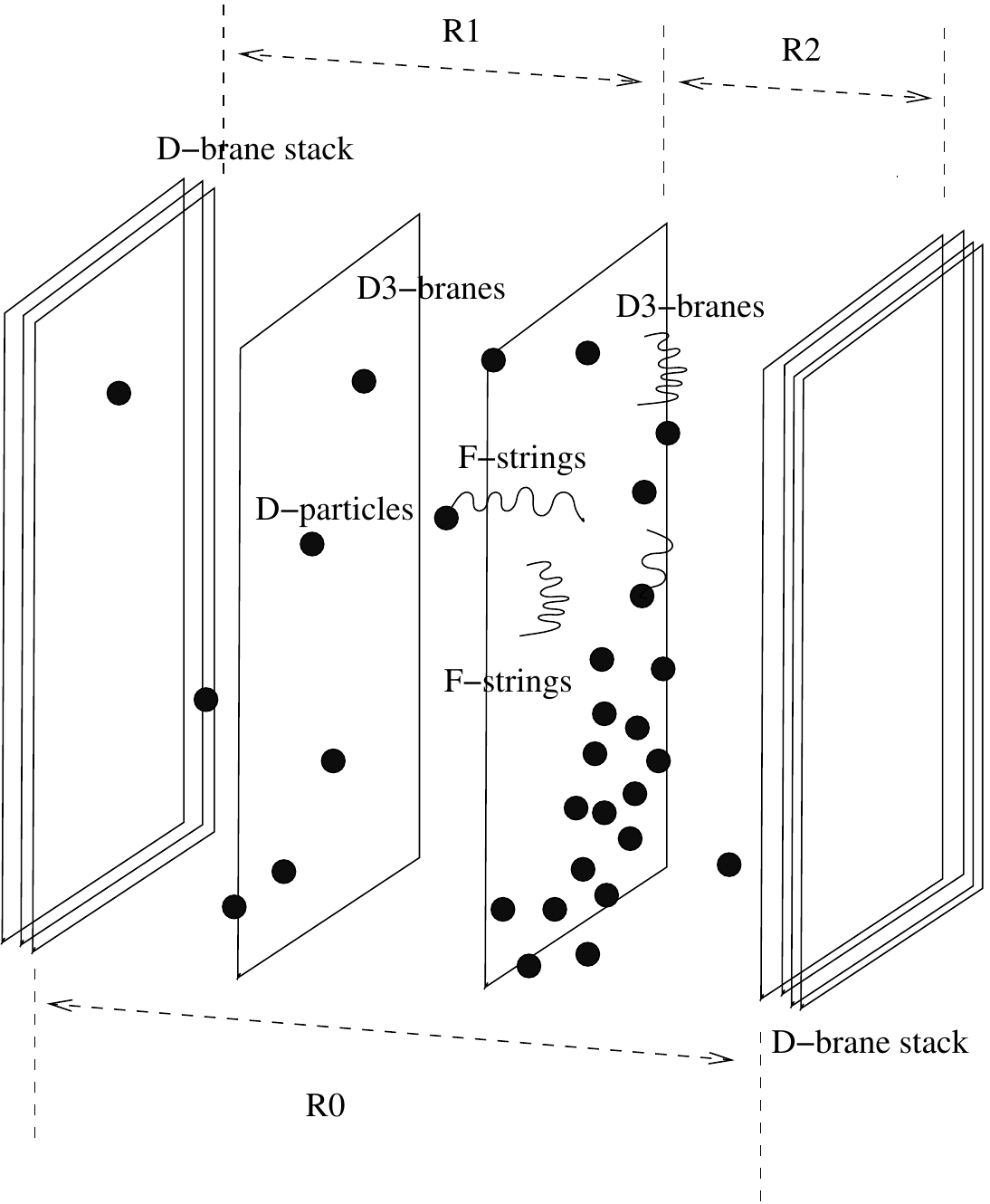} \caption{Schematic representation of a prototype D-particle space-time foam model~\cite{dfoam}, 
consisting of two stacks of higher-dimensional D-branes, 
attached to orientifold planes, which, due to their special reflective properties, provide a natural compactification of the bulk
dimension. The bulk is punctured by D0-branes (D-particles).
Our ``world'' is one of the brane Universes, after appropriate compactification to three large spatial dimensions (D3 branes). Open fundamental (F-)strings live on the D3-brane world, representing matter excitations of the Standard Model. Matter can interact topologically 
with the D-particle defects in the foam, {\it e.g.}, through capture and splitting of the open string by the defect,
re-emission of the open string, and recoil of the D-particle. In each such process there are distortions of the neighboring space-time.}
\label{fig:dfoam}
\end{figure}

Another field-theoretic  context where  an asymptotic  space-time with
the form (\ref{defspace}) emerges is that of the so-called D-particle
``foam''. In  this scenario, the Universe  is modelled by a  brane world
propagating  in a  higher  dimensional bulk  space-time, punctured  by
stochastically  fluctuating  D0-branes  (or  D-particles)~\cite{dfoam}
(\emph{cf.}     fig.~\ref{fig:dfoam}).     In the case of a 
four-dimensional brane world,
fundamental (F-)  strings propagate on the brane, representing 
matter  and/or  radiation  excitations   of  the  observable  Universe
(e.g. Standard  Model (SM) fields).
These strings may be captured by the defect,  
causing the attachment of at  least one of  its ends  to the
D-particle.  Subsequently,  the  open  string is  re-emitted,  and  the
D-particle  recoils,   a  process  which  involves   the  creation  of
fundamental  strings stretched  between the  D-particle and  the brane
Universe.  In  the presence  of  an  ensemble of  quantum  fluctuating
D-particles (D-foam) such a process is repeated several times, and one
essentially has to average  over statistically significant populations
of the D-particles in order  to describe, at an effective (low-energy)
field theory  level, the propagation  of an open string  excitation in
such a ``medium''.  The recoil of the D-particle defect
implies  a distortion  of  the  neighbouring space-time 
by an  amount
proportional to the  momentum transfer exchanged in the process. 

Assuming  a locally flat  space-time, the
corresponding metric distortion in the rest frame of the D-particle
can be calculated by noting~\cite{recoil} the similarity of the problem
with that of an open string (representing SM excitations on the brane world) in an external ``electric'' field~\cite{susskind} of intensity $u_i = \frac{\Delta p_i}{M_s} g_s$, 
where $u_i$ is the recoil velocity of the D-particle on the D3-brane world, along its $i$-th spatial large dimension, as seen by a cosmic 
observer who is at rest with respect to the brane universe, and 
$\Delta p_i$ is the momentum transfer of the matter excitation in that frame, with $M_s/g_s$ the D-particle mass, $M_s$ the string mass scale, and $g_s <1 $ the string coupling. 
In this  frame, the distorted metric ``felt'' by the open string
is then given by (in spherical polar coordinates, assuming - without loss of generality- recoil along the radial direction)~\cite{recoil}
\be\label{recmetric}
g_{\mu\nu} = \Big[\begin{array}{c} (1 - |u_r|^2)\, \eta_{\mu\nu}, \quad \mu, \nu = 0, r \\
\eta_{\mu\nu}, \quad \quad \quad \quad  \qquad \mu, \nu = \theta, \phi \end{array}
\ee
with $\eta_{\mu\nu}$ the Minkowski metric with signature $(-, +, +, +)$. It should be noted that there is an underlying \emph{non-commutative} geometry between temporal and spatial coordinates in this case~\cite{susskind,recoil}, 
\be\label{noncommut}
\Big[ t, x_i \Big] = i \frac{u_i}{1 - |u_i|^2}~,
\ee
and hence the effects of D-particle recoil are expected to lead to physically non-trivial results, such as a refractive index for photons propagating in this background~\cite{mitsou}, or, as we discuss next, an angular \emph{surplus} (negative deficit) in the space-time felt by SM particles.

Indeed, upon averaging over ensembles of D-particles, using the stochastic relations 
\be\label{stoch}
\ll u_i \gg =0, \, \quad \ll u_i \, u_j \,u_k \gg =0, \,  \quad \ll u_i \, u_j \gg = \sigma^2 \, \delta_{ij}~,  \quad i,j=1,2,3~, \quad \sigma^2 \ll 1~,
\ee
it is then straightforward to write the induced metric element (\ref{recmetric}) as
\be\label{average}
\ll ds^2 \gg = -(1 - \sigma^2) dt^2 +(1 - \sigma^2 ) dr^2 + r^2 (d\theta^2 + {\rm sin}^2\theta \, d\phi^2)~,  \qquad \sigma^2 \ll 1~, 
\ee
which, upon a trivial rescaling of the time coordinate by the factor $(1 - \sigma^2)$ implies a metric of the type (\ref{defspace}), 
\bea\label{average2}
\ll ds^2 \gg &=&- dt^{\prime\,2} + dr^{\prime\, 2} + \frac{1}{1-\sigma^2} \, r^{\prime \,2} (d\theta^2 + {\rm sin}^2\theta \, d\phi^2) \nonumber \\
 &\simeq & -dt^{\prime\,2} + dr^{\prime\, 2} + (1 +\sigma^2) \, r^{\prime \,2} (d\theta^2 + {\rm sin}^2\theta \, d\phi^2)~,  \qquad \sigma^2 \ll 1~,
\eea
where the corresponding \emph{surplus}  (negative deficit in this case) angle (\ref{deficit}) is given by the stochastic D-particle
recoil velocities variance $\sigma^2 \ll 1$, which is a characteristic property of the foam. Experimentally,
for a dilute foam (which is the physically expected situation) it may be possible in principle (depending on the magnitude of $\sigma^2$)
to falsify the phenomenon, given that there should be enhanced scattering patterns in a small angular region in the forward directions, 
where the scattering (of photons in this case) from a distant astrophysical object will be enhanced compared to the one expected in the absence of D-foam.
The phenomenon would manifest itself as an excess of photons compared to the expected flux in the absence of the foam. Such phenomena could be combined
with the induced anisotropies in superheavy dark matter scattering by the D-foam, examined in \cite{mp}.
Disentangling such phenomena in cosmological searches from standard dark matter searches is an open issue, which will not be the subject of the current article.

In all the above situations, the scattering of standard model particles off such defects will exhibit the scattering lensing phenomenon described above,
which we now proceed to analyze in the following sections.

\section{Quantum Scattering on space-time defects\label{sec:lensing}}

In this section we review in detail 
the analysis of \cite{mazur} for the case of massless scalar neutral particles, in a space-time with an angular deficit (\ref{deficit}) of the form (\ref{defspace}).

To that end, we employ the notation of \cite{mazur},
which has the advantage of being general and not specific to the details of the underlying microscopic model,
and rewrite the space-time element  
as~\footnote{For definiteness, we restrict ourselves below to the deficit angle case, for which $0 < b \le 1$. For the case of surplus angular defect, as happens in the example of D-foam (\ref{average2}), the parameter $ b > 1$, but, apart from this, our subsequent analysis and conclusions on the physical phenomenon of lensing hold in that case as well.}
\be\label{spacedef}
ds^2  = -d{t}^2 + d{r}^2 +   b^2\, {r}^2 \Big( d\theta^2 + {\rm sin}^2 \theta \, d\phi ^2 \Big)~, \quad b^2 \le 1~, \quad b \in {\tt R}~.
\ee
The only point we make about the deficit parameter $b$ is that it is close to 1, so that appropriate perturbative expansions are valid.  In the static space-time (\ref{spacedef}), one can  parametrize the wave corresponding to the scalar field $\Phi (\vec x,t) = e^{i \omega t} \, \Psi (r,\theta, \phi)$, 
where $\omega$ is the energy of the massless field, $\omega =|\vec k| \equiv k$, with $k$ the spatial momentum. 
The (Klein-Gordon) equation of motion $g^{\mu\nu} \nabla_\mu \, \partial_\nu \, \Phi = 0$ (with $\nabla$ a gravitational covariant derivative), then, reduces to a Helmholtz-type equation for $\Psi (r,\theta, \phi)$ ~\cite{mazur}
\be\label{helm}
\Delta \, \Psi = \omega^2 \, \Psi~, \quad \Delta \equiv -\frac{1}{r^2} \, \frac{\partial}{\partial \, r} \Big( r^2 \, \frac{\partial}{\partial \, r} \Big) - \frac{{\tt L}^2}{b^2 \, r^2}
\ee
with $\tt L$ the Laplacian of a unit sphere, corresponding to the ``angular momentum'' operator. 

The spherical symmetry of the problem allows one to employ as an orthonormal basis the Legendre polynomials $P_\ell ({\rm cos}\theta) $,  $\ell =0, 1, 2, \dots $, which satisfy~\cite{wu,abram}
\be\label{ang}
{\tt L}^2 \, P_\ell ({\rm cos}\theta)  = \ell (\ell + 1) \, P_\ell ({\rm cos}\theta) , \quad \ell \in {\tt N}_0~.
\ee
In terms of this basis, the function $\Psi (r, \theta, \phi)$ can be expanded as~\cite{mazur}
\be\label{expansion}
\Psi (r, \theta, \phi) = \sum_{\ell =0}^{\infty} \, c_\ell \, R_\ell (r) \, P_\ell ({\rm cos}\theta)
\ee
Before proceeding, we consider it instructive to list some properties of the Legendre polynomials, which we shall use in our analysis below. 
A particular property of the Legendre polynomials is the relationship of a special sum of them to the Dirac delta function~\cite{abram}
\be\label{delta}
\delta (y-x) = \sum_{\ell=0}^{\infty} \, \left(\ell + \frac{1}{2} \right)\, P_\ell (y) \, P_\ell (x) ~, -1 \le \, y \, \le 1, \quad 1 \le \, x \, \le 1~.
\ee
Another useful quantity is their generating function
\be\label{gener}
\frac{1}{\sqrt{1 - 2x\, t\, + t^2}} = \sum_{\ell=0}^{\infty} t^\ell \, P_\ell (x) ~, 
\ee
where the variable $t$ can be complexified, $ t \to w \in {\tt C}$, through analytic continuation.  From (\ref{gener}) we also obtain, by Taylor expanding the left hand side, that $P_0(x)=1, \, P_1(x)=x$.
The following ``normalization'' of the Legendre polynomials will be adopted 
\be\label{norm}
P_\ell (1) = 1~, \quad \ell \in {\tt N}_0~,
\ee
which can be achieved by an appropriate scaling, since both the orthogonality property and the differential equation defining the Legendre polynomials~\cite{abram} are independent of scaling. The result (\ref{norm}), when used in conjunction with (\ref{delta}), implies for $y=1$, $x={\rm cos}\theta$ (\emph{cf.} (\ref{expansion})),
\be\label{delta2}
\delta (1- {\rm cos}\theta) = \sum_{\ell=0}^{\infty} \, \left(\ell + \frac{1}{2} \right)\, P_\ell ({\rm cos}\theta) ~,
\ee

Returning to the expansion (\ref{expansion}), and using (\ref{helm}) and (\ref{ang}), we obtain 
\be\label{interR}
R^{\prime\prime}_\ell + \frac{2}{r} \, R^\prime_\ell + \left(\omega^2 - \frac{\ell (\ell + 1)}{b^2 \, r^2} \right)\!\! R_\ell =0~.
\ee
On writing  $R_\ell(r)=r^{-1/2} \, {\mathcal G}_\ell (r)$, and noticing that in (\ref{interR}) one can scale $\omega r \to y$, and treat $y$  as the differential equation variable, one finally obtains from (\ref{interR}) the following second order differential equation
 \be\label{RG}
 r^2 \, {\mathcal G}_\ell^{\prime\prime} + r\, {\mathcal G}_\ell^\prime + \left(r^2 \, \omega^2 - \left[\frac{\ell (\ell +1 )}{b^2} + \frac{1}{4} \right]
 \right)\!{\mathcal G}_\ell =0~.
\ee
The above equation admits as solution~\cite{mazur} spherical Bessel functions of the first kind ${J}_{\nu (\ell)}$, of order $\nu(\ell)$~\cite{abram},
\begin{eqnarray}\label{bessel}
{\mathcal G}_{\ell}(r) &=& {J}_{\nu(\ell)} (\omega \, r) ~, \nonumber \\ \nu(\ell) &=& b^{-1} \left[ \left(\ell + \frac{1}{2}\right)^2  - \frac{1-b^2}{4}\right]^{1/2} = \left[ \frac{\ell (\ell + 1)}{b^2} + \frac{1}{4}\right]^{1/2} \nonumber \\
&=& \ell + \frac{1}{2} - \frac{2}{\pi}\, \delta_\ell, \qquad \delta_\ell :=  \frac{\pi}{2} \left[ \left(\ell + \frac{1}{2}\right) -
  b^{-1} \sqrt{\left(\ell + \frac{1}{2}\right)^2 - \frac{1-b^2}{4} }\,\,\right] ~, 
\end{eqnarray}
where we restricted ourselves to the finite solution as $r \to 0$, which is the one with physical significance in our case, providing a smooth connection with the no-defect limit~\footnote{For other boundary conditions, see \cite{obc}.}.
As we will see, the quantity $\delta_\ell$ will be identified with the phase shift caused by the scattering of the particle off the defect. 

To discuss (quantum) scattering, we now write the function $\Psi (r) $ in (\ref{expansion}) as a sum of an incoming ($\Psi_{\rm in}$) and a scattered ($\Psi_{\rm sc}$) wave,
\be\label{psi}
\Psi = \Psi_{\rm in} + \Psi_{\rm sc},
\ee
with 
\be\label{psiin}
\Psi_{\rm in} = e^{i \omega \, r \, {\rm cos}\theta} \,,
\ee
 assuming, for concreteness, propagation of the incident wave along the $z$ axis, and 
the scattering solution at $r \to \infty$
\be\label{scatter}
\Psi_{\rm sc}  (r \to \infty) \sim  \frac{1}{r} \, f(\theta) e^{i \omega \, r} \,, 
\ee
where $f(\theta)$ is the scattering amplitude in our quantum mechanical formulation~\cite{wu}. We also impose that $\Psi_{\rm sc} \to 0$ when $b \to 1$,
which specifies uniquely $f(\theta)$~\cite{mazur}.

From (\ref{scatter}) and (\ref{expansion}) we obtain in the asymptotic region  $r \to \infty$
\be\label{Rinf}
R_\ell (r \to \infty) =  \lim\limits_{r \to \infty} r^{-1/2} G_\ell (r) = \lim\limits_{r \to \infty} r^{-1/2} \, {J}_{\nu(\ell)} (\omega \, r) \simeq  \sqrt{\frac{2}{\pi \omega}} \, r^{-1} \, {\rm cos}\left(\omega\, r - \frac{\pi \, \nu(\ell)}{2} - \frac{\pi}{4}\right)~, 
\ee
where in the last equality we have used for $r \to \infty$
the asymptotic form of the Bessel function $J_{\nu(\ell)}(\omega \, r )$, which is regular at the origin. 
Using (\ref{bessel}) we can express the argument of the cosine function in (\ref{Rinf}) in terms of the phase shift $\delta_\ell$, thus finally obtaining for the function $\Psi(r \to 0,\theta,\phi)$ in (\ref{expansion}) 
\begin{eqnarray}\label{asymptBessel}
\Psi (r \to \infty, \theta,\phi ) &\simeq &  \frac{1}{r} \,\sum_{\ell=0}^{\infty} c_\ell \, 
\sqrt{\frac{2}{\pi \omega}} \, {\rm cos}\left(\omega\, r - \frac{\pi \, (\ell +1)}{2} + \delta_\ell\right) \, P_\ell ({\rm cos}\theta) = \nonumber \\
& = &  \frac{1}{r} \, \sum_{\ell=0}^{\infty} c_\ell \,  \sqrt{\frac{1}{2\, \pi \omega}} \, \left(e^{i(\omega\, r - \frac{\pi \, (\ell +1)}{2} + \delta_\ell)} + e^{-i(\omega\, r - \frac{\pi \, (\ell +1)}{2} + \delta_\ell)} \right) \, P_\ell ({\rm cos}\theta)~.
\end{eqnarray}
We now express the exponential $e^{i\omega r {\rm cos}\theta}$ in terms of appropriate sums of the Bessel functions $J_n (x)$ as~\cite{abram}
\bea\label{expbess}
e^{i\omega \, r \, {\rm cos}\theta} &=& \sqrt{2 \, \pi}\, \sum_{\ell =0}^{\infty}\, \left(\ell + \frac{1}{2}\right)\, i^\ell \,
\frac{J_{\ell+\frac{1}{2}}(\omega\, r)}{(\omega \, r)^{1/2}}\, P_\ell ({\rm cos}\theta)  
\nonumber \\ 
&\stackrel{r \to \infty} \simeq &  \frac{1}{\omega \, r} \, \sum_{\ell =0}^{\infty}\, \left(\ell + \frac{1}{2}\right)\, i^\ell \,  \left(e^{i(\omega\, r - \frac{\pi \, (\ell +1)}{2})} + e^{-i(\omega\, r - \frac{\pi \, (\ell +1)}{2} )} \right) \, P_\ell ({\rm cos}\theta)~,
\eea
where again in the last line we used the asymptotic form of the Bessel function $J_n(x)$ for $x \to \infty$. 

Substituting (\ref{expbess}) into (\ref{psi}), taking into account (\ref{psiin}), (\ref{scatter}), equating (\ref{psi}) with (\ref{asymptBessel}) for $r \to \infty$, and finally equating the respective 
coefficients of $e^{\pm \, i\omega \, r}$, we obtain the expressions for the coeffcients $c_\ell$ in (\ref{asymptBessel}) and the scattering amplitude $f(\theta)$
\be\label{cl}
c_\ell = \sqrt{\frac{2\,\pi}{\omega}} \, \left( \ell + \frac{1}{2} \right) \, i^\ell \, e^{i\delta_\ell}~,
\ee
and
\bea\label{amplitude}
f(\theta) = -\frac{i}{\omega}  \, \sum_{\ell=0}^{\infty}  \left(\ell + \frac{1}{2}\right) \, \left(e^{2i\delta_\ell} -1 \right) \, P_\ell ({\rm cos}\theta)~, 
\eea
or, equivalently, 
\be\label{ampldelta}
f(\theta) = \frac{1}{\omega}  \, \sum_{\ell=0}^{\infty}  \left(2\, \ell + 1\right) \, e^{i\,\delta_\ell} \, {\rm sin}\delta_\ell \, P_\ell ({\rm cos}\theta)~, 
\ee
where $\delta_\ell$ is given in (\ref{bessel}). 

We next proceed to expand $\delta_\ell$ as a power series of the small variable $1- b^2 \simeq 2\,\alpha, \, b \to 1^-$ (\emph{i.e.} small deficit angle (\ref{deficit})), which is relevant for our physically interesting cases discussed in the previous section.
In particular, keeping only the leading order approximation, and setting $\alpha := 1 - b^{-1} \to 0$,
we obtain 
\be
\label{shift}
\delta_\ell \stackrel{b \to 1^-} \simeq  \frac{\pi}{2} \, \alpha \,  \left(\ell + \frac{1}{2}\right)
+ \frac{\pi \, (1 - b^2)}{16\, b \,\left(\ell + \frac{1}{2}\right) } \,.
\ee

We observe from (\ref{amplitude}) and (\ref{shift}) that $f(\theta) \to 0$ as $b \to 1$, since in that case $\delta_\ell \to 0$, in agreement with our boundary condition $\Psi_{\rm sc} \to 0$ when $b \to 1$. Using (\ref{delta2}) we may write the scattering amplitude (\ref{amplitude}) as
\be\label{amplitude2}
f(\theta) = -\frac{i}{\omega} \, \sum_{\ell=0}^{\infty}  \left(\ell + \frac{1}{2}\right) \, e^{2i\delta_\ell} \, P_\ell ({\rm cos}\theta) + \frac{i}{\omega}\, \delta(1 - {\rm cos}\theta)~.
\ee
The presence of the $\delta$-function on the right-hand side of (\ref{amplitude2}), which was omitted in the initial analysis of \cite{mazur}, is crucial for ensuring 
that in the absence of the deficit, {\it i.e.} $ b \to 1$ and $\delta_\ell \to 0$ in (\ref{shift}), the amplitude $f(\theta) \to 0$,
and, therefore, any potential phenomenon disappears as the Minkowski space-time is recovered. 

To discuss further the consequence of the deficit $b \ne 1$ in the scattering off a defect, one might be tempted to expand the $e^{i\delta_\ell}$ in powers of a small $\alpha \to 0$ (\ref{shift}), which is the case of physical interest, keeping only leading terms in the expansion. However, this is not correct, given that $\alpha \ell $ can be much greater than unity for sufficiently large $\ell$. Hence it is appropriate to only partially expand the exponent in $e^{2i\delta_\ell}$ by writing
\be\label{exponentdl}
e^{2i\, \delta_\ell}  \, \stackrel{\alpha \ll 1} \simeq \,  e^{i\, \pi \, \alpha \, \left(\ell + \frac{1}{2}\right)}
\left[1 + i\, \frac{\pi\, (1-b^2)}{8\, b\, \, \left(\ell + \frac{1}{2}\right)} \right]~,
\ee
which implies that the amplitude (\ref{amplitude2}) can be written as
\bea\label{amplitude2b}
f(\theta) &=& -\frac{i}{\omega} \,  \sum_{\ell=0}^{\infty} \, \left( \ell + \frac{1}{2}\right)  \, (e^{i\pi\, \alpha})^{\ell + \frac{1}{2}} \, P_\ell ({\rm cos}\theta)  \, +  \,
\frac{\pi \, (1-b^2)}{8\, b\, \omega}\,  \sum_{\ell=0}^{\infty} \, (e^{i\pi\, \alpha})^{\ell + \frac{1}{2}}  \, P_\ell ({\rm cos}\theta)
\nonumber\\
&& +  \frac{i}{\omega}\, \delta(1 - {\rm cos}\theta)~.
\eea
Writing 
\be
\sum_{\ell} \, \left(\ell + \frac{1}{2} \right)  \,  (e^{i\pi\, \alpha})^{\ell  + \frac{1}{2}  \,   }\, P_\ell ({\rm cos}\theta) = -\frac{i}{\pi} \, \frac {d}{d\, \alpha}\, \sum_\ell \, (e^{i\pi\, \alpha})^{\ell + \frac{1}{2}}\, P_\ell ({\rm cos}\theta)~,
\ee
and 
making use of the generating function of the Legendre polynomials (\ref{gener}), with $x={\rm cos}\theta$ and $t=e^{i\pi\,\alpha}$, implying that 
\be\label{identity}
\sum_{\ell=0}^{\infty} \, (e^{i\pi\, \alpha})^{\ell + \frac{1}{2}} \, P_\ell ({\rm cos}\theta) = \frac{1}{\sqrt{2\,\Big({\rm cos\pi \alpha - {\rm cos}\theta\Big)}}}~,
\ee
one readily obtains from (\ref{amplitude2b})
\be\label{amplitude3}
f(\theta) = -\frac{1}{\, 2\sqrt{2}\, \omega} \,  \frac{{\rm sin}\pi\alpha}{\Big({\rm cos}\pi\alpha - {\rm cos}\theta\Big)^{\frac{3}{2}}} + \frac{1}{\omega}\, \frac{\pi (1-b^2)}{8\, \sqrt{2}\,b} \, \frac{1}{\Big({\rm cos}\pi\alpha - {\rm cos}\theta\Big)^{\frac{1}{2}}} +  \frac{i}{\omega}\, \delta(1 - {\rm cos}\theta)~.
\ee
 It is clear from the above expression that the scattering amplitude  diverges for the special value of the scattering angle $\theta=\theta^\star = |\pi \alpha|$.  This is the essence of the lensing phenomenon discussed in \cite{mazur}. In view of the singular behaviour of (\ref{amplitude3}) we coin this phenomenon \emph{singular lensing}.

 The fact that the dominant contribution to the scattering amplitude occurs when \mbox{$\theta = \pm \pi\alpha$} may also be understood by noting that,  for large $\ell$,
 the asympotic form of the Legendre polynomials is given by~\cite{wu}
\be\label{asympt}
P_\ell ({\rm cos}\theta) \stackrel{\ell \gg 1}\simeq \sqrt{\frac{2}{\pi\, \ell \, {\rm sin}\theta}} \, \left(1 - \frac{1}{8\ell \theta} {\rm cos}\left[\left(\ell + \frac{1}{2}\right)\theta + \frac{\pi}{4}\right]\right)\,
{\rm sin}\left[\left(\ell + \frac{1}{2}\right)\theta + \frac{\pi}{4}\right]
\ee
Substituting the above expression into (\ref{amplitude}), we obtain for the scattering amplitude for large $\ell$ and $\theta \ne 0$, such that $\ell \, \theta \gg 1$:
\be\label{ampllarge}
f(\theta) \stackrel{\ell \theta \gg 1}\simeq -\frac{1}{2\,\omega}\, \sum \, \sqrt{\frac{2 \ell}{\pi\, {\rm sin}\theta}} \, e^{2i\delta_\ell} \, \Big(e^{i(\ell+\frac{1}{2})\theta -i \frac{\pi}{4}} -
e^{-i(\ell+\frac{1}{2})\theta + i\frac{\pi}{4}}\Big),
\ee
from which we observe that, due to the oscillatory behaviour of the exponentials with $\ell \theta$, the dominant contributions in (\ref{ampllarge}) come from those $\ell$ for which the 
exponents $2\delta_\ell \pm \ell \theta$ do not vary much with $\ell$, that is~\cite{wu} 
\be\label{deltatheta}
2\frac{d \delta_\ell}{d  \, \ell}  \pm \theta \simeq 0 \quad \Rightarrow \quad \delta_\ell =\pm\frac{1}{2}\ell \, \theta ~.
 \ee 
In our case, to leading order in large $\ell$, we have that $\delta_\ell \stackrel{\ell \gg 1}\simeq  \frac{1}{2} \ell \pi\alpha $, from which it follows immediately that a dominant contribution to the amplitude should come when $\theta$ is in the vicinity of $\pm \alpha$, in agreement with (\ref{amplitude3}). However, notice that in our case this contribution is formally divergent, as can also be seen by replacing the large-$\ell$ summation in (\ref{ampllarge}) by a continuous integral over $\ell$. Indeed, expanding $e^{2i\delta_\ell} = {\rm cos}(2\delta_\ell) + i \, {\rm sin}(2\delta_\ell)$
in (\ref{ampllarge}),  with $2\delta_\ell \sim \pi\alpha $ as $\ell \to \infty$, leads to integral structures of the form (up to irrelevant $\ell$-independent factors that do not affect our arguments)
\bea\label{pnea}
f(\theta) & \stackrel{\ell \to \infty}\propto & {\mathcal I}_1 + i {\mathcal I}_2, \nonumber \\
{\mathcal I}_1 &=& \int_{\ell \to \infty}\!\!\!\! d\ell \, \sqrt{\ell} \, {\rm cos}(\pi\alpha\, \ell) \, {\rm sin}(\theta \, \ell)  
\nonumber \\
&\simeq & \frac{1}{2} \, \left[\sqrt{\frac{\pi}{2}} \,
 \frac{{\rm Fr_C} (\sqrt{\ell z_{-} })}{\phi_{-}^{3/2}} + 
 \frac{ {\rm Fr_C} (\sqrt{\ell z_{+} })}{\phi_{+}^{3/2}}
 - 2 \sqrt{\ell}\, \frac{\theta \, {\rm cos}(\theta\ell) \, {\rm cos}(\pi\alpha \, \ell) + \pi\alpha \, {\rm sin}(\theta\, \ell) \, {\rm sin}(\pi\alpha \, \ell)}
 {\phi_{+} \phi_{-}}~\right]
\nonumber \\
&\stackrel{\ell \to \infty} \simeq &  {\mathcal O}\left(\sqrt{\ell} \, {\rm sin}(\ell\,\phi_{\pm}) \right)\,,
 \nonumber \\
{\mathcal I}_2 &=& \int_{\ell \to \infty} \!\!\!\! d\ell \, \sqrt{\ell} \, {\rm sin}(\pi\alpha\, \ell) \, {\rm sin}(\theta \, \ell) \nonumber \\
&\simeq & \frac{1}{2} \, \left[- \sqrt{\frac{\pi}{2}} \,\frac{{\rm Fr_S} (\sqrt{\ell z_{-} })}{\phi_{-}^{3/2}} + 
\frac{ {\rm Fr_S} (\sqrt{\ell z_{+} })}{\phi_{+}^{3/2}}
+ \sqrt{\ell}\, \left(\frac{{\rm sin} \phi_{-}}{\phi_{-}} -  \frac{{\rm sin} \phi_{+}}{\phi_{+}}\right)\right]~
\nonumber \\
& \stackrel{\ell \to \infty} \simeq &  {\mathcal O}\left(\sqrt{\ell} \, {\rm sin}(\ell\,\phi_{\pm}) \right)\,,
\eea
where $\phi_{\pm} :=\theta \pm \pi\alpha $,  $z_{\pm} := (2/\pi)\phi_{\pm}$, 
and ${\rm Fr_S} (x) = \int_0^x \, dt \, {\rm sin}(t^2) $, 
${\rm Fr_C}(x) = \int_0^x \, dt \, {\rm cos}(t^2) $ denote the Fresnel integrals~\cite{abram}, which, in the limit $x \to \infty$, behave as 
$ {\rm Fr_S}(x)  = {\rm Fr_C}(x) \stackrel{x \to \infty}\simeq 
\sqrt{\frac{\pi}{2}} \Big(\frac{{\rm sign}(x)}{2} + {\mathcal O}(\frac{1}{x}) \,\Big)$. 
The integrations in the above limit have been performed with {\it Mathematica}. 

In the limit $\theta = \pi\alpha \ne 0$ we obtain for the above integrals
\bea\label{pea}
{\mathcal I}_1 &\stackrel{\ell \to \infty,\, \pi\alpha=\theta}\simeq &-4\pi(\pi\alpha)^2 \, \sqrt{\ell}\, {\rm cos}(2\,\pi\alpha \, \ell) + {\mathcal O}\Big(\sqrt{\frac{1}{\ell}}\Big)~, \nonumber \\
{\mathcal I}_2 &\stackrel{\ell \to \infty,\, \pi\alpha= \theta}\simeq &\frac{1}{3}\, \ell^{\,3/2} + {\mathcal O}\Big(\sqrt{\frac{1}{\ell}}\Big)~, \nonumber \\
\eea
The reader should compare the distinct ways the integrals diverge as $\ell \to \infty$ between the two cases (\ref{pnea}), (\ref{pea}). In the case (\ref{pea}), where $\pi\alpha = \theta$, the leading divergence in $f(\theta)$ is of order 
$\ell^{3/2}$, which is 
much stronger than that in the case (\ref{pnea}) $\theta\ne \pi\alpha$, where it is suppressed by infinitely rapidly oscillating trigonometric functions, being of the form $\sqrt{\ell} \, {\rm sin}(\ell \, (\theta \pm \pi\alpha))$. 
In fact, the latter terms can be resummed, when the full series for all $\ell$ is considered, yielding (\ref{amplitude3}), which is finite for $0 < \theta\ne -\pi\alpha $ (the $\delta$-function term vanishes for $\theta \ne 0$). 
This is, once again, the singular lensing phenomenon found earlier. 
  
For completeness we note~\cite{pusc} that, upon assuming a non-zero $\pi\alpha \ne 0$, the amplitude acquires different values for the cases $\theta < \pi \alpha $ and $\theta > \pi\alpha$:
\bea\label{amplitude4}
f(\theta)\,{\Big|_{\theta < \pi\alpha}}  &=& -\frac{i}{\, 2\sqrt{2}\, \omega} \,  \frac{1}{\left({\rm cos}\theta - \cos\pi\alpha \right)^{\frac{1}{2}}} \left[
\frac{\sin\pi\alpha}{{\rm cos}\theta - \cos\pi\alpha} + \frac{\pi (1-b^2)}{4\, \,b} \right]  +  \frac{i}{\omega}\, \delta(1 - {\rm cos}\theta)~, \nonumber \\
f(\theta)\,{\Big|_{\theta > \pi\alpha}}  &=& \frac{1}{\, 2\sqrt{2}\, \omega} \,  \frac{1}{\left({\rm cos}\pi\alpha - {\rm cos}\theta\right)^{\frac{1}{2}}} \left[ - 
\frac{{\rm sin}\pi\alpha}{{\rm cos}\pi\alpha  - {\rm cos}\theta} + \frac{\pi (1-b^2)}{4\, \,b} \, \right]~,
\eea
where we took into account that the $\delta$-function vanishes in the case $\theta > \pi\alpha \ne 0$. 

 For future use we remark that, for $\pi\alpha \ne 0$,  the relations (\ref{amplitude4}) imply
\be\label{imf}
{\rm Im}f(\theta=0) =  -\frac{1}{\, 2\sqrt{2}\, \omega} \,  \frac{1}{\left(1 - {\rm cos}\pi\alpha \right)^{\frac{1}{2}}} \left[
\frac{{\rm sin}\pi\alpha}{\left(1 - {\rm cos}\pi\alpha\right)} + \frac{\pi (1-b^2)}{4\, \,b} \right]  +  \frac{1}{\omega}\, \delta(0)~,
\ee
where $\delta (0)$ should be understood as a term in need of proper regularization, to be discussed below. 

\section{Recovering the ``no-defect'' limit \label{sec:nodeflimit}}

The subject of this section is related with a consistency check of our approach, namely 
with demonstrating that the $f(\theta)$ in (\ref{amplitude3}) satisfies 
the boundary condition $f(\theta) \to 0$, as $b \to 1$, which was imposed on $\Psi_{sc}$ (\ref{scatter}), and ought to specify uniquely $f(\theta)$, as already mentioned.
The transition to the ``no-defect'' limit yields automatically the correct (vanishin) result  
when one defines $f(\theta)$ by means of summation over Legendre polynomials, (\ref{amplitude}), from which follows trivially that when $b \to 1$, and thus $\delta_\ell \to 0$ (\ref{shift}) for each partial wave $\ell$ , then $f(\theta) \to 0$. It is instructive, however, to verify this explicitly at the level of (\ref{amplitude3}),
as the latter involved several algebraic manipulations of the various infinite sums entering in (\ref{amplitude}).

The first subtlety in (\ref{amplitude3}) is the range of $\theta$. For any finite $\theta \ne 0$, the $\delta$-function term 
$\delta(1 -{\rm cos}\theta) \to 0$, and in this case,  we observe from (\ref{amplitude}) that, for $b \to 1$ (hence, $\alpha \to 0$ as well), the condition $f(\theta \ne 0, b \to 1)$ is satisfied. 
The subtle point is the limit $\theta \to 0$, for which the $\delta$-function $\lim\limits_{\theta \to 0} \delta(1- {\rm cos}\theta) \to \delta (0)$ is formally infinite and needs regularization.  

Setting formally $\theta =0 $ in the first of (\ref{amplitude4}), and defining the approach of $\pi\alpha \to 0$ by replacing
\be\label{prescr}
\pi\, \alpha  \to \pi\alpha + \epsilon \to 0, \quad \epsilon \rightarrow 0^+, \quad \epsilon \gg |\pi\alpha| \quad {\rm as} \, \alpha \to  0^-,
\ee
with $\epsilon $ an awlays positive quantity independent of $\pi\alpha$, 
we have for the leading divergent term of the  first line (\ref{amplitude4}) (the first one on the right-hand-side) as $b \to 1$:
\bea\label{amplitudelimit}
f(\theta = 0)  &\stackrel{b \to 1} \simeq &-\frac{i}{\omega} \frac{\pi\alpha + \epsilon}{\Big( (\pi\alpha)^2 + \epsilon^2 \Big)^{3/2}} + \frac{i}{\omega} \delta(0) + \dots 
 \stackrel{0 \leftarrow |\pi\alpha | \ll \epsilon \to 0^+}\simeq -\frac{i}{\omega}  \, \frac{\epsilon}{(\epsilon)^3} + \frac{i}{\omega} \delta(0) + \dots 
\nonumber \\
&\stackrel{0 \leftarrow |\pi\alpha | \ll \epsilon \to 0^+}\simeq  & - \frac{i}{\omega} \, \frac{1}{\epsilon^2} + \frac{i}{\omega} \delta(0) + \dots~, \quad \epsilon \to 0^+, \, \alpha \to 0^-, \quad \epsilon \gg |\pi\alpha |,
\eea
with the ellipses indicating subleading finite terms, stemming from the $(1-b^2)$-terms in (\ref{amplitude4}) as $b \to 1^-$
\be\label{finite}
\Big[ f(\theta=0) \Big]_{{\rm Finite~Parts}}\, \stackrel{b\to 1^-}= \, -\frac{i\, {\rm sign}(\pi\alpha + \epsilon)}{4 \, \omega} = \,-\frac{i\, \epsilon}{4 \, \omega}~,\quad  \epsilon \to 0^+, \, \pi \alpha=0~,
\ee
given that $\pi\, (1-b^2)/b  \stackrel{\,\,\,b\to 1^-} \simeq -2 \pi \alpha \to -2\pi\alpha - 2 \epsilon$ in the non-defect limit, due to our prescription (\ref{prescr}). 

It should be stressed that the prescription (\ref{prescr}) guarantees that, in the no-defect
limit $\theta=|\pi\alpha| \to 0^+$, one can always cancel the singular and negative $\epsilon$-dependent
terms in (\ref{amplitudelimit}) by the non-negative term involving the $\delta$-function distribution, in a way independent of the sign
of the deficit $\pi\alpha$, namely,  by defining the \emph{regularized} singular limit $\delta(0)$
such that it cancels \emph{both} the leading divergent and finite  ((\ref{finite})) terms in (\ref{amplitudelimit}) as $b \to 1^-$, $\epsilon \to 0^+$,
\be\label{reguldelta}
\delta(0) \, \stackrel{b \to 1^-}{=:} \, \frac{1}{\epsilon^2} + \frac{1}{4} ~, \quad   \epsilon \to 0^+, \quad \pi\alpha =0.
\ee
This is a self-consistent prescription, in agreement with the boundary condition that \mbox{$f(\theta) \to 0$}, as $b \to 1$, which is respected in (\ref{amplitude}).  
The prescription (\ref{reguldelta}), to leading order as $\epsilon \to 0^+$, can also be viewed as the following regularization of the $\delta(0)$
\be\label{deltaregul2}
\delta(0) =  \sum_{\ell =0}^{\infty} \Big(\ell + \frac{1}{2}\Big) \, e^{i \epsilon (\ell + \frac{1}{2})} ~, \quad \epsilon \to 0^+~,
\ee
with $\epsilon \to 0^+$ defined as in (\ref{prescr}), which makes manifest the vanishing of the scattering amplitude $f(\theta)$  (\ref{amplitude2b}) in the no-defect limit. In fact, for a generic scattering problem with a phase shift $\delta_\ell$, we may define a regularized version of (\ref{amplitude2}), using (\ref{deltaregul2}), as follows
\be\label{amplitudereg}
f(\theta) = -\frac{i}{\omega} \, \sum_{\ell=0}^{\infty}  \left(\ell + \frac{1}{2}\right) \, \Big( e^{2i\delta_\ell + i\,\epsilon (\ell+\frac{1}{2})} \, P_\ell ({\rm cos}\theta) -   e^{i \, \epsilon (\ell + \frac{1}{2})}\Big)
~, \quad \epsilon \to 0^+~.
\ee
This definition was missed in the previous literature,  where the behaviour of the scattering amplitude in the no-defect limit was incompletely addressed.

\section{Differential Cross section and lensing \label{sec:pheno}}

In this section we proceed to discuss some phenomenological aspects of the production at particle  colliders 
of defects that lead to space-times with a conical deficit solid angle.

\begin{figure}[ht!]
\includegraphics[width=0.6\columnwidth]{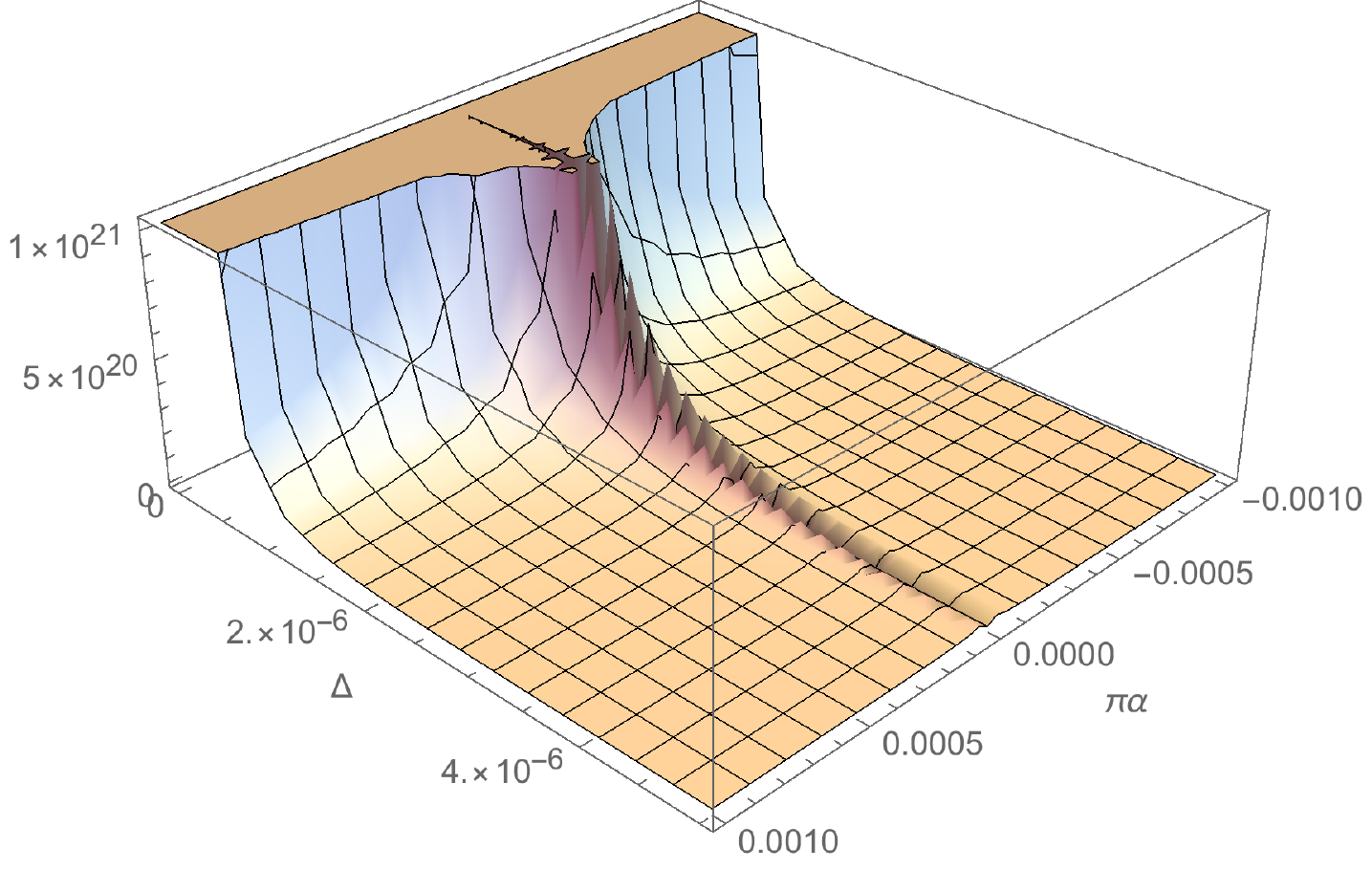}  \hfill
\includegraphics[width=0.6\columnwidth]{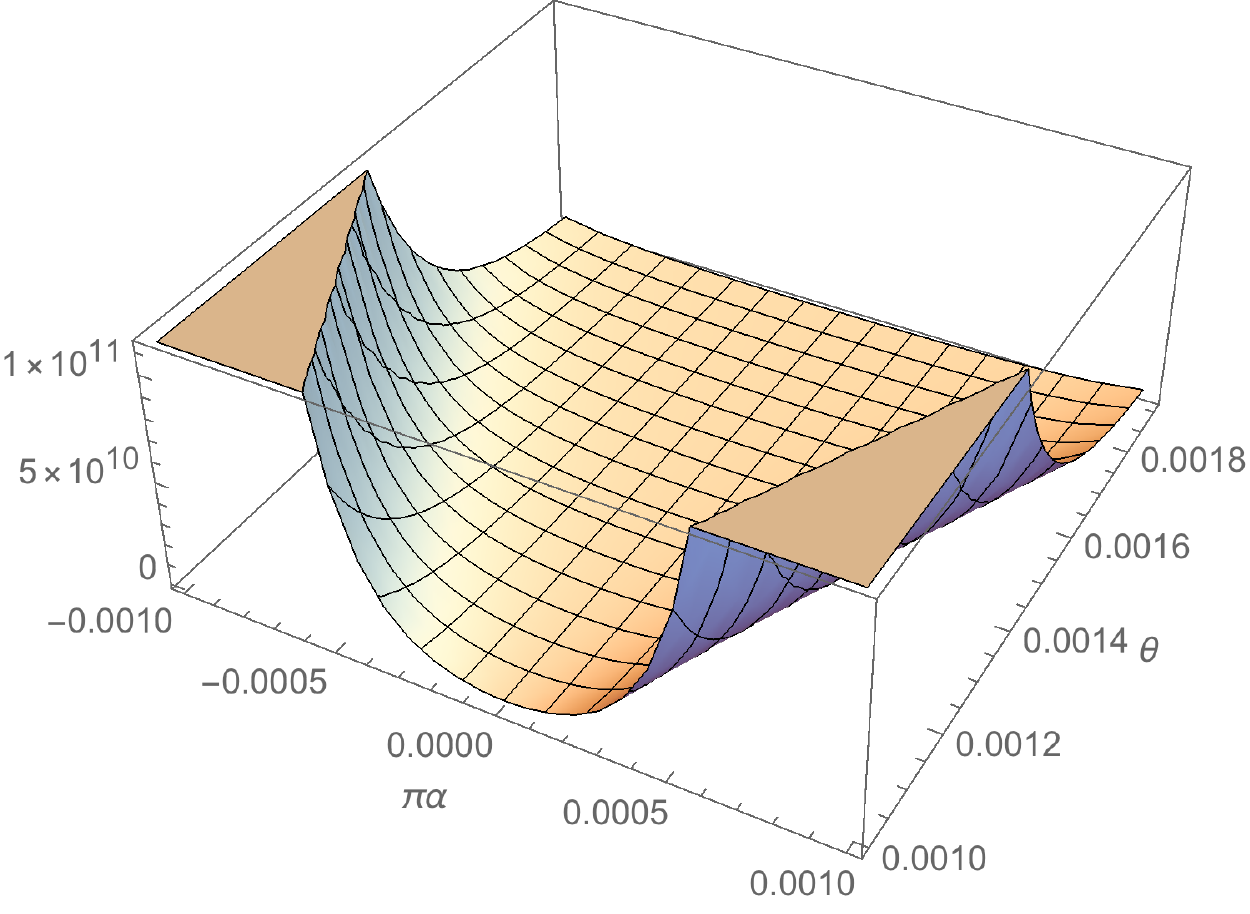} 
\caption{Three dimensional plots of the differential cross section (\ref{analyticdc}), as a function of either $\Delta=\theta - |\pi\alpha| >0$  and $\pi\alpha$ (upper panel), or $\theta$ and $\pi\alpha$ (lower panel), for the case $\theta \ge |\pi\alpha| \ne 0$. The lensing of particles when $\theta \to |\pi\alpha | \ne 0$ is evident. We deliberately kept both signs of $\pi\alpha$ (although in the concrete cases studied in this section $\alpha \le 0$), to demonstrate a branch cut at $\pi\alpha=0$, which defines the no deficit limit, for which the cross section vanishes.}
\label{fig:cs}
\end{figure}

The differential cross section of the scattering of massless scalar fields off the defect is given by
\be\label{diffcross}
\frac{d \sigma }{d \Omega}  = |f(\theta)|^2 ~, \qquad d\Omega  = {\rm sin}\theta\, d\theta \, d\phi~,
\ee
where $\Omega$ is the three-dimensional solid angle, expressed in spherical coordinates. From (\ref{amplitude3}) one observes that, for $\theta=0$ and $b \ne 1$ ($\alpha \ne 0$), the differential cross section (\ref{diffcross}) is singular due to the $\delta$-function term. This is an important aspect of the \emph{presence} of the defect, yielding a focus point of the scattered particles in the forward direction. 
In addition to the $\theta =0$ case, one also has a formal divergence of the amplitude (\ref{amplitude2}), and hence of the differential cross section (\ref{diffcross}),  for the case $ |\pi\alpha| = \theta \ne 0$, which was the effect discussed in \cite{mazur}, and reproduced in various other occasions in \cite{others1,others2,lousto,pusc}. In that case, we obtain from (\ref{diffcross}) and (\ref{amplitude4}):
\bea\label{analyticdc}
\frac{d \sigma }{d \Omega} & \stackrel{\theta \ge -\pi\alpha} =& \frac{1}{8\, \omega^2} \, \frac{{\rm sin}^2\pi\alpha}{\left({\rm cos}\pi\alpha - {\rm cos}\theta \right)^3} \, \left[ 1 - \frac{\pi\, (1-b^2)}{4\,b} \frac{\left({\rm cos}\pi\alpha - {\rm cos}\theta \right) }{{\rm sin}\pi\alpha}\, \right]^2  \nonumber \\
&=& \frac{1}{64\, \omega^2} \,  \frac{{\rm sin}^2\pi\alpha}{\left({\rm sin}(\frac{\Delta}{2}) \, {\rm sin}(\frac{\Delta}{2} + |\pi\alpha|) \right)^3} \, \left[1 - \frac{\pi\, (1-b^2)}{2\,b}\, 
\frac{\left({\rm sin}(\frac{\Delta}{2}) \, {\rm sin}(\frac{\Delta}{2} + |\pi\alpha|) \right)}{{\rm sin}\pi\alpha}\, \right]^2~,\nonumber \\
\eea
where in the second line we have expressed the result in terms of the (non-negative) parameter~\cite{pusc} $\Delta~\equiv \theta - |\pi\alpha| \ge 0$, using the simple trigonometric relation
${\rm cos}\pi\alpha - {\rm cos}\theta = 2\, \Big({\rm sin}(\frac{\Delta}{2}) \, {\rm sin}(\frac{\Delta}{2} + |\pi\alpha|) \Big)$. 
This allows the physical effects of the limit $\theta \to |\pi\alpha| \ne 0$ (\emph{i.e.} when $0 < \Delta \ll |\pi\alpha| $) to be more easily visualised.
The differential cross section (\ref{analyticdc}) is plotted in Figure \ref{fig:cs}.

Indeed, the leading behaviour of the differential cross section (\ref{analyticdc}), as $\theta \to |\pi\alpha| \ne 0$ ($0 < \Delta \ll |\pi\alpha|$), is
\be\label{leaddiffcross}
\frac{d \sigma }{d \Omega}  \stackrel{\theta \to |\pi\alpha| \ne 0} \simeq \frac{{\rm sin}^2\pi\alpha}{8\, \omega^2 \, \Big({\rm cos}\pi\alpha  - {\rm cos}\theta\Big)^3} 
 \stackrel{0< \Delta \ll |\pi\alpha|}\simeq \,  \frac{1}{64\, \omega^2 \, |{\rm sin}\pi\alpha |\, {\rm sin}^3(\frac{\Delta}{2})} ~.
\ee
which diverges for $\Delta \to 0$, giving rise is the lensing phenomenon~\cite{mazur}. 

Of course, in practice, this divergence will be regulated by the experimental angular resolution $\theta_{\rm res}$, which imposes a natural cut-off $\Delta \ge \theta_{\rm res}$ in the above expressions. This would tame the apparent divergence in the value of the differential cross section 
in Fig.~\ref{fig:cs}; the maximum value attained will be $\frac{d\sigma}{d\Omega}|_{\rm max} = \frac{d\sigma (\Delta=\theta_{\rm res})}{d\Omega}$. In fact, $\theta_{\rm res}$ acts as a regulator also of the formally singular quantity $\delta(0)$, as discussed at the end of the next section.

We next remark that by considering the case $\pi\alpha \ll 1$, which is of physical interest as becomes clear from our discussion in section \ref{sec:models}, 
then in the region of scattering angles such that
$\pi\alpha \ll \Delta$,  we have from (\ref{analyticdc}) a suppressed differential cross section
\be\label{analyticdc2}
\frac{d \sigma }{d \Omega}  \stackrel{|\pi\alpha| \ll \Delta\, , \,|\pi\alpha | \ll 1}\simeq \frac{(\pi\alpha)^2}{64\, \omega^2\,
  {\rm sin}^6(\frac{\Delta}{2})} \left[1 + {\rm sin}^2\left(\frac{\Delta}{2}\right) \right]^2~,
\ee
where we took into account that in the case $|\pi \alpha | \ll 1$ we can employ 
the approximation $\pi (1-b^2)/{2 b} \simeq - \pi\alpha >0$.

\section{The optical theorem \label{sec:appendix}}

In this section we address certain subtleties related with the way that
the optical theorem is realized in the case of the process considered above.

In its text-book formulation,  the optical theorem relates the total (elastic) cross-section with the
imaginary part of the forward scattering amplitude, as 
\be\label{optical}
\sigma_{\rm tot} = \frac{4\pi}{\omega} \, {\rm Im}\, f(0) \,,
\ee
where, in our case, $f(0)$ is given by $f_{\theta < \pi\alpha}(0)$ in (\ref{amplitude3}) for $\pi\alpha \ne 0$,
and its imaginary part is given in (\ref{imf}). According to the standard lore,   
the validity of the theorem follows from unitarity of the scattering matrix, or,
equivalently, from the conservation of probability at the level of the wave function.
Even though in the presence of gravitational interactions the notion of unitarity may be tricky, 
for the problem at hand, namely for scattering far away from the defect core, 
unitarity in the standard sense is expected to be valid, and hence, the optical theorem should hold. 

Formally the validity of the theorem follows from the expression (\ref{ampldelta}) for the scattering amplitude as an infinite sum of partial waves, and the integral of the differential cross section (\ref{diffcross}) over the solid angle $d\Omega$ in three-space (below, the $*$ denotes complex conjugation and we set from now on  $x \equiv {\rm cos}\theta$):
\bea\label{optical2}
\sigma_{\rm tot} &=& \int_0^{2\pi} \, d\phi \, \int_{0}^{\pi} {\rm sin}\theta \, d\theta \, f({\rm cos}\theta) \, f^{*}({\rm cos}\theta) =2\pi \, \int_{-1}^1 \, dx\, |f(x)|^2
 \nonumber \\
 &=& 2\pi \, \int_{-1}^1 \, dx \, \frac{1}{\omega^2} \, \sum_{\ell,m} \, (2\ell+1)\, (2m+1) \, {\rm sin}\delta_\ell \, {\rm sin}\delta_m \, P_\ell (x) \, P_m(x)  
= \frac{4\pi}{\omega^2} \, \sum_{\ell} (2\ell+1) {\rm sin}^2\delta_\ell  \nonumber \\
\eea
where in the last equality  we used the orthogonality relation of the Legendre polynomials~\cite{abram}:
\be\label{ortho}
\int_{-1}^1 \, dx \, P_\ell (x) \, P_m(x) = \frac{2}{2m + 1} \, \delta_{\ell \, m}
\ee
with $\delta_{\ell \, m}$ the Kronecker delta. Then, the optical theorem (\ref{optical})  follows immediately from (\ref{optical2}) on account of (\ref{ampldelta}), upon recalling the normalization (\ref{norm}) of the Legendre polynomials.

As a non-trivial consistency check of our approximations, we 
derive next the total cross section by explicitly integrating the approximate differential cross section (\ref{diffcross}) over the above range of $(\theta,\phi)$; evidently,   
the validity of the optical theorem (\ref{optical}) should not be taken for granted when dealing with this truncated expression.
The integrated version of (\ref{diffcross}) is given by  
\bea\label{totalfromdiff}
\sigma_{\rm tot} = \int d\Omega |f(\theta)|^2 =  2\pi \Big( \int_{-1}^{{\rm cos}\pi\alpha} \, dx \, |f_{\theta > |\pi\alpha|}(x)|^2  +
\int^{1}_{{\rm cos}\pi\alpha} \, dx \, |f_{\theta < |\pi\alpha|}(x)|^2  \Big) ~,
\eea
where we have carried out the trivial integration over the azimuthal angle $\phi$, 
and the amplitudes in the integrands are given by (\ref{amplitude3}). 
The problem is that the above integration involves singular limits, which have to be carefully regularized. In doing so, we will postulate the validity of the optical theorem (\ref{optical}),
which will serve as our guiding principle  in determining the exact regularization procedure.

The pertinent integrals have the structure (after appropriate change of integration variable $x \to x/{\rm cos}\pi\alpha$)
\be\label{integrals}
I_1= \int_{-\tfrac{1}{{\rm cos}\pi\alpha}}^{1} \, dy \, \big(1 - y)^{-d}~, \quad d=1,2 \quad  ; \quad 
I_2= \int^{\tfrac{1}{{\rm cos}\pi\alpha}}_{1} \, dy \, \big(y - 1)^{-d}~, \quad d=1,2,3~,
\ee
and the divergences in question are 
associated with the upper (lower) integration limit in the first (second) integral. Therefore, 
a careful cutting-off procedure is required, with a cut-off $\tilde \epsilon \to 0^+$, that we proceed to discuss next. 
The regularization should also be such that, for $\alpha \to 0$ (no defects), the cross  section should vanish identically, as discussed earlier. 

We next mention some additional points that will be essential for the computation of (\ref{totalfromdiff}).
In our analysis we encounter terms involving the square of the Dirac $\delta$-function, of the form
\be\label{deltasquare}
{\mathcal A} \equiv \frac{2\pi}{\omega^2} \, \Big( \int^{1}_{{\rm cos}\pi\alpha} \, dx\, \delta^2(1-x)  \Big) = 
\frac{2\pi}{\omega^2} \, \int_{-1}^{1} \, dx\, \delta^2(1-x)~,  
\ee
where in the last equality we extended the integration by adding an identically zero term (as the Dirac $\delta$-function vanishes in the region 
$-{\rm cos}\pi\alpha < x < {\rm cos}\pi\alpha $ for $\pi\alpha \ne 0$ we are considering here). 
Making use of (\ref{delta2}), (\ref{ortho}), and (\ref{norm}), we can write (\ref{deltasquare}) as 
\bea\label{deltasquarefinal}
{\mathcal A} &=&  \frac{2\pi}{\omega^2} \, \int_{-1}^{1} \, dx\, 
 \sum_{\ell, m} \, \left(\ell + \frac{1}{2} \right)\, \left(m + \frac{1}{2} \right) 
 P_\ell (x) \, P_m (x)  \nonumber \\ &=&  \frac{2\pi}{\omega^2} \, \sum_{\ell} \, \left(\ell + \frac{1}{2} \right) = \frac{2\pi}{\omega^2} \, \delta(0)~.
\eea
We also encounter integrals of the form  
\be\label{heaviside}
\int_{{\rm cos}\pi\alpha}^{1}\, dx \, \delta(1 - x) \, \frac{1}{(x - {\rm cos}\pi\alpha)^c} = \Big(\frac{1}{1 - {\rm cos}\pi\alpha}\Big)^c\,\Theta(0) ~, \quad c \in {\tt R}~.
\ee
with the convention for the Heaviside $\Theta(x)$ function at $x=0$, 
\be\label{norm2}
\Theta (0) =1,
\ee
for the problem at hand, where the functions involved are right-continuous, in view of (\ref{amplitude4}).
Then, after some elementary integrations, 
we easily derive from (\ref{totalfromdiff})
\bea\label{tfdiff}
\sigma_{\rm tot} &=& \frac{\pi \, {\rm tan}^2\pi\alpha}{4\,  \omega^2} \, \frac{1}{\tilde \epsilon^2}  - \frac{\pi \,(1 + {\rm cos}^2\pi\alpha)}{4\, \omega^2 \, {\rm sin}^2\pi\alpha}  \nonumber \\
&+& \frac{\pi^3 \,(1-b^2)^2}{64\, b^2 \, \omega^2} \, {\rm ln}\Big(\frac{1- {\rm cos}\pi\alpha}{1 +{\rm cos} \pi\alpha}\Big)   -
\frac{\pi^2 \, (1-b^2)}{4\, b\, \omega^2} \, \frac{{\rm cos}\pi\alpha}{{\rm sin}\pi\alpha} 
\nonumber \\
&+& \frac{4\pi}{\omega^2} \Big(-\frac{1}{\, 2\sqrt{2}} \,  \frac{1}{\Big(1 - {\rm cos}\pi\alpha \Big)^{\frac{1}{2}}} \Big[
\frac{{\rm sin}\pi\alpha}{\Big(1 - {\rm cos}\pi\alpha\Big)} + \frac{\pi (1-b^2)}{4\, \,b} \Big]  +  \frac{1}{2}\, \delta(0) \Big)~, 
\eea
with the cut-off $\tilde \epsilon \to 0^+$ has been introduced. 
As a consistency check, we note that the right-hand-side of (\ref{tfdiff}) is positive definite. 

The reader should notice that the last line of (\ref{tfdiff}) would constitute the part of the total cross section if the coefficient of the singular term $\delta(0)$
were unity. In other words, adding and subtracting $\frac{2\pi}{\omega^2}\, \delta(0)$, we obtain from (\ref{tfdiff})
\be\label{final}
\sigma_{\rm tot} = \frac{4\pi}{\omega}\, {\rm Im}f(0) + \frac{\pi}{\omega^2} \, {\mathcal E}
\ee
where 
\bea\label{extra}
{\mathcal E} := \frac{{\rm tan}^2\pi\alpha}{4\, \tilde \epsilon^2}  - \frac{(1 + {\rm cos}^2\pi\alpha)}{4\, {\rm sin}^2\pi\alpha}  
+ \frac{\pi^2 \,(1-b^2)^2}{64\, b^2} \, {\rm ln}\Big(\frac{1- {\rm cos}\pi\alpha}{1 +{\rm cos} \pi\alpha}\Big)   -
\frac{\pi \, (1-b^2)}{4\, b} \, \frac{{\rm cos}\pi\alpha}{{\rm sin}\pi\alpha} - 2\, \delta (0)\,.  \nonumber \\
\eea
To restore this, we should postulate a choice of the cut-off  $\tilde \epsilon \to 0^+$ such that $\mathcal E = 0$  for any  $\pi\alpha \ne 0$. This can be easily enforced
by absorbing the 
$\pi\alpha$-dependent terms in (\ref{extra}) in the definition of the cutoff. 
In doing so we employ the $\alpha$-independent regularization of $\delta (0) = \frac{1}{\epsilon^2} + \frac{1}{4}, \, \epsilon \to 0^+$,  given in (\ref{reguldelta}).

This regularization guarantees the optical theorem (\ref{optical}) and is consistent with the vanishing of the cross section (and the amplitude $f(\theta)$) in the no-defect limit $\pi\alpha \to 0$, as it is compatible with the regulated $\delta(0)$ (\ref{reguldelta}). In fact, in that limit, one should consider the replacement  $\pi\alpha \to \pi\alpha + \epsilon$ (\emph{cf.}  (\ref{prescr})), as $\pi\alpha \to 0^-$, with $ |\pi\alpha | \ll \epsilon \to 0^+$. In such a case we have:
\be\label{relations}
\frac{1}{\tilde \epsilon^2} \sim \frac{10}{\epsilon^4} +  \dots   \to \infty, \quad {\rm as} \quad \epsilon \to 0^+, \quad  0^+ \leftarrow |\pi\alpha| \ll \epsilon,
\ee
where the ellipses indicate (irrelevant) subleading terms. 

 We finally point out that the infinities in the differential cross section discussed above
are the result of considering quantum-mechanical instead of quantum-field-theoretic scattering, including gravitons; the latter would 
include effects of back reaction onto the (curved) spacetime, ignored in the current analysis, 
which are expected to smoothen out the singularities in (\ref{leaddiffcross}), while preserving the characteristic  
enhancement in angular regions where $\theta \sim \pi\alpha$. 

A final, but important comment is due at this point. In practice, the $\delta (0)$ appearing in (\ref{tfdiff}) or (\ref{optical}) is replaced by the value of a $\delta$-function distribution at the experimental angular resolution $\theta_{\rm res}$, which is considered to be small. In order to have a
phenomenon, one must have $|\theta_{\rm res}| < |\pi\alpha|$, which prompts one to 
use the analogue of (\ref{reguldelta}) for representing ``experimentally'' the quantity $\delta(0)$,
\be\label{resolution}
\delta(0) \rightarrow \delta_{\rm expt}(\theta_{\rm res}^2) \simeq \frac{1}{\theta_{\rm res}^2} + \frac{1}{4}~, \qquad \theta_{\rm res} < |\pi\alpha |~,
\ee
given that  $f(\theta)$ should vanish when $\theta < \theta_{\rm res}$.
This is because, for scattering angles $0 < \theta \le \theta_{\rm res}$, one cannot distinguish experimentally the 
forward scattered particles from the unscattered incident beam.
The relation (\ref{resolution}) should be used when discussing the potential phenomenology related to this effect, which was done in section \ref{sec:pheno}.

This completes our discussion on the regularized cross sections, which, as we have seen, is a subtle and delicate issue.

\section{Conclusions and Outlook \label{sec:concl}}

In the present work we have revisited the problem of particle scattering off a global defect, which is known to induce 
a space-time with an angular deficit or surplus. For concreteness we have focused on the deficit case, but our results may be straightforwardly extended  
 to a space-time with an angular surplus, such as those found in the D-foam systems. 
 Our analysis demonstrates that the effect of particle lensing  is mathematically robust, surviving a proper regularization of the Legendre polynomial series. 
 Within this framework, we have verified the disappearance  of the effect in the no-defect limit, and the validity of the optical 
 theorem for the total elastic cross-section.  Even though we explicitly studied the spin 0 case, the generalization to fermions~\cite{others1} and gauge-bosons 
 may be carried out in a similar manner. 
  
  The phenomenon has potentially wide applications due to the variety of physical systems that may produce it. 
  The important point to notice is that our analysis has been restricted to electrically neutral particles, because in the presence of 
  electromagnetic (Coulomb) interactions  of charged matter, the effect, which is essentially gravitational in origin, would  be strongly suppressed. 
  Should global defects be produced in colliders, only neutral particles will be lensed due to this effect. Such a lensing may manifest itself
 through the excess of photons (either primarily produced or stemming from the decays of other neutral particles)
 in regions of the detectors corresponding to the ring-like structures associated with the phenomenon.   
  
  We now remark that, if the defects are solitonic in nature, as in \cite{vilenkin,sarben}, their production in colliders is expected to be strongly suppressed~\cite{drukier}. Nonetheless, as already mentioned at the end of section \ref{sec:anttensor}, enhanced production of structured defects may be foreseen in the presence of strong magnetic fields and/or at high temperatures, as happens in the environment of a neutron star or in heavy ion collisions. This is the result of a thermal analogue of Schwinger pair production~\cite{rajantie}, provided of course that the deficit is present in such situations, in the sense that the temperature 
 has not restored the broken symmetry. 

 Cosmological applications of this phenomenon are also very interesting~\cite{mazur}, since in this case it will manifest 
 itself as ring-like structures of cosmic photons (predominantly cosmic microwave radiation) in the sky. 
 In models of space-time D-foam~\cite{dfoam,recoil,mp}, which can be used as alternatives to dark matter~\cite{sakell}, such structures may 
 provide a natural explanation for potentially observed photon excesses,
 which would be conventionally interpreted as being due to the 
 annihilation of dark matter particles.    Moreover, in view of the similarity of the global monopole space-time with that of cosmic strings, searches for the lensing phenomenon 
 can be included in the current efforts~\cite{hindmarsh} to locate such defects in the Universe. Let us finally note that cosmic neutrinos will also exhibit the lensing effect, which may in principle lead to enhanced signals in detectors.
  
\section*{Acknowledgements} 
 
The authors would like to acknowledge discussions with V. A.~Mitsou.
 N.E.M. wishes to thank the University of Valencia and IFIC for a Distinguished Visiting Professorship, during the tenure of which this work has been completed. The work of N.E.M. is also supported in part by the Science and Technology Facilities Council (STFC), UK, under the research grant ST/P000258/1. The research of J.P. 
is supported by the Spanish Ministry of Education and Science (MEYC) under FPA2014-53631-C2-1-P and SEV-2014-0398,
and Generalitat Valenciana under grant Prometeo II/2014/066.


\begin{thebibliography}{99} 

\bibitem{dirac} P.~A.~M.~Dirac,
  Phys.\ Rev.\  {\bf 74}, 817 (1948).
  doi:10.1103/PhysRev.74.817


\bibitem{shir}
Y.~M.~Shnir,
  ``Magnetic Monopoles,''
  doi:10.1007/3-540-29082-6

\bibitem{gravmon} P.~Breitenlohner, P.~Forgacs and D.~Maison,
  Nucl.\ Phys.\ B {\bf 383}, 357 (1992).
  doi:10.1016/0550-3213(92)90682-2;
  Nucl.\ Phys.\ B {\bf 442}, 126 (1995)
  doi:10.1016/S0550-3213(95)00100-X
  [gr-qc/9412039].

\bibitem{poincare} H. Poincar\'e, 
\emph{Remarques sur une exp`erience de M. Birkeland}, Comptes Rendus Acad.
Sci. \textbf{123}, 530 (1896),

\bibitem{birkeland} K. Birkeland, Elec. Rev. \textbf{38}, 752, 782 (1896); Arch. Sci. Phys. Nat. \textbf{1}, 497 (1896);
Elektrotechknik (Wien) \textbf{14}, 448, 475 (1896).

\bibitem{hpmono} G.~'t Hooft,
  Nucl.\ Phys.\ B {\bf 79}, 276 (1974).
  doi:10.1016/0550-3213(74)90486-6;
  A.~M.~Polyakov,
  JETP Lett.\  {\bf 20}, 194 (1974)
  [Pisma Zh.\ Eksp.\ Teor.\ Fiz.\  {\bf 20}, 430 (1974)].

\bibitem{monscat} See, \emph{e.g.}: D.~G.~Boulware, L.~S.~Brown, R.~N.~Cahn, S.~D.~Ellis and C.~k.~Lee,
  Phys.\ Rev.\ D {\bf 14}, 2708 (1976).
  doi:10.1103/PhysRevD.14.2708;
K.~A.~Milton,
  Rept.\ Prog.\ Phys.\  {\bf 69}, 1637 (2006)
  doi:10.1088/0034-4885/69/6/R02
  [hep-ex/0602040] and references therein.
P.~Rossi,
  Phys.\ Rept.\  {\bf 86}, 317 (1982).
  doi:10.1016/0370-1573(82)90081-3, and references therein.

\bibitem{vilenkin} M.~Barriola and A.~Vilenkin,
  Phys.\ Rev.\ Lett.\  {\bf 63}, 341 (1989).
  doi:10.1103/PhysRevLett.63.341
  
  \bibitem{sarben} N.~E.~Mavromatos and S.~Sarkar,
  Phys.\ Rev.\ D {\bf 95}, no. 10, 104025 (2017)
  doi:10.1103/PhysRevD.95.104025
  [arXiv:1607.01315 [hep-th]].


\bibitem{dfoam} J.~R.~Ellis, N.~E.~Mavromatos and M.~Westmuckett,
  Phys.\ Rev.\ D {\bf 70}, 044036 (2004)
  doi:10.1103/PhysRevD.70.044036
  [gr-qc/0405066];
  {\it ibid.} \ D {\bf 71}, 106006 (2005)
  doi:10.1103/PhysRevD.71.106006
  [gr-qc/0501060].
J.~R.~Ellis, N.~E.~Mavromatos and D.~V.~Nanopoulos,
  Phys.\ Rev.\ D {\bf 62} (2000) 084019
  doi:10.1103/PhysRevD.62.084019
  [gr-qc/0006004].
  
   
  \bibitem{recoil}  N.~E.~Mavromatos,
  Found.\ Phys.\  {\bf 40}, 917 (2010)
  doi:10.1007/s10701-009-9372-z
  [arXiv:0906.2712 [hep-th]]. 

  
    
\bibitem{mazur} P.~O.~Mazur and J.~Papavassiliou,
  Phys.\ Rev.\ D {\bf 44}, 1317 (1991).
  doi:10.1103/PhysRevD.44.1317;
  
  \bibitem{sakell} T.~Elghozi, N.~E.~Mavromatos, M.~Sakellariadou and M.~F.~Yusaf,
  JCAP {\bf 1602}, no. 02, 060 (2016)
  doi:10.1088/1475-7516/2016/02/060
  [arXiv:1512.03331 [hep-th]].

  
  \bibitem{others1}  H.~Ren,
  Phys.\ Lett.\ B {\bf 325}, 149 (1994)
  doi:10.1016/0370-2693(94)90085-X
  [hep-th/9312074];
  
\bibitem{others2} E.~R.~Bezerra de Mello and C.~Furtado,
  Phys.\ Rev.\ D {\bf 56}, 1345 (1997).
  doi:10.1103/PhysRevD.56.1345
A.~A.~Roderigues Sobreira and E.~R.~Bezerra de Mello,
  Grav.\ Cosmol.\  {\bf 5}, 177 (1999)
  [hep-th/9809212];

\bibitem{cosmic} A.~Vilenkin,
  Phys.\ Rept.\  {\bf 121}, 263 (1985).
  doi:10.1016/0370-1573(85)90033-X


\bibitem{lousto} C.~O.~Lousto,
  Class.\ Quant.\ Grav.\  {\bf 9}, 2417 (1992).
  doi:10.1088/0264-9381/9/11/008.




\bibitem{hindmarsh} A.~Lopez-Eiguren, J.~Lizarraga, M.~Hindmarsh and J.~Urrestilla,
  JCAP {\bf 1707}, no. 07, 026 (2017)
  doi:10.1088/1475-7516/2017/07/026
  [arXiv:1705.04154 [astro-ph.CO]].

\bibitem{atlasmon} G.~Aad {\it et al.} [ATLAS Collaboration],
  Phys.\ Rev.\ D {\bf 93}, no. 5, 052009 (2016)
  doi:10.1103/PhysRevD.93.052009
  [arXiv:1509.08059 [hep-ex]];
  Phys.\ Rev.\ Lett.\  {\bf 109}, 261803 (2012)
  doi:10.1103/PhysRevLett.109.261803
  [arXiv:1207.6411 [hep-ex]].
  
  
\bibitem{moedal} 
  B.~Acharya {\it et al.} [MoEDAL Collaboration],
  Int.\ J.\ Mod.\ Phys.\ A {\bf 29}, 1430050 (2014)
  doi:10.1142/S0217751X14300506
  [arXiv:1405.7662 [hep-ph]]. For a review of updated recent results and prospects see, \emph{eg.} 
  V.~A.~Mitsou [MoEDAL Collaboration],
  PoS CORFU {\bf 2016}, 028 (2017).


\bibitem{pusc} W.~Puszkarz,
  gr-qc/9509060.

\bibitem{debate} For a partial list of references on the (still open) issue of the stability of the global monopole, see: A.~S.~Goldhaber,
  Phys.\ Rev.\ Lett.\  {\bf 63}, 2158 (1989).
  doi:10.1103/PhysRevLett.63.2158;
In the original suggestion of Goldhaber that global monopoles are not stable against ``angular'' collapse, there is an ongoing debate on this issue; for a partial list of references  see:
S.~H.~Rhie and D.~P.~Bennett,
  Phys.\ Rev.\ Lett.\  {\bf 67}, 1173 (1991).
  doi:10.1103/PhysRevLett.67.1173;
L.~Perivolaropoulos,
  Nucl.\ Phys.\ B {\bf 375}, 665 (1992).
  doi:10.1016/0550-3213(92)90115-R;
G.~W.~Gibbons, M.~E.~Ortiz, F.~Ruiz Ruiz and T.~M.~Samols,
  Nucl.\ Phys.\ B {\bf 385}, 127 (1992)
  doi:10.1016/0550-3213(92)90097-U
  [hep-th/9203023];
M.~Hindmarsh,
  Nucl.\ Phys.\ B {\bf 392}, 461 (1993)
  doi:10.1016/0550-3213(93)90681-E
  [hep-ph/9206229];
G.~Arreaga, I.~Cho and J.~Guven,
  Phys.\ Rev.\ D {\bf 62}, 043520 (2000)
  doi:10.1103/PhysRevD.62.043520
  [gr-qc/0001078];
A.~Achucarro and J.~Urrestilla,
  Phys.\ Rev.\ Lett.\  {\bf 85}, 3091 (2000)
  doi:10.1103/PhysRevLett.85.3091
  [hep-ph/0003145];
  R.~Gregory and C.~Santos,
  Class.\ Quant.\ Grav.\  {\bf 20}, 21 (2003)
  doi:10.1088/0264-9381/20/1/302
  [hep-th/0208037];
  E.~R.~Bezerra de Mello,
  Phys.\ Rev.\ D {\bf 68}, 088702 (2003)
  doi:10.1103/PhysRevD.68.088702
  [hep-th/0304029];
  S.~B.~Gudnason and J.~Evslin,
  Phys.\ Rev.\ D {\bf 92}, no. 4, 045044 (2015)
  doi:10.1103/PhysRevD.92.045044
  [arXiv:1507.03400 [hep-th]].

 \bibitem{negative} D.~Harari and C.~Lousto,
  Phys.\ Rev.\ D {\bf 42}, 2626 (1990).
  doi:10.1103/PhysRevD.42.2626

\bibitem{bronnikov} K.~A.~Bronnikov, B.~E.~Meierovich and E.~R.~Podolyak,
  J.\ Exp.\ Theor.\ Phys.\  {\bf 95}, 392 (2002)
  [Zh.\ Eksp.\ Teor.\ Fiz.\  {\bf 122}, 459 (2002)]
  doi:10.1134/1.1513811
  [gr-qc/0212091].


\bibitem{drukier} 
  A.~K.~Drukier and S.~Nussinov,
  Phys.\ Rev.\ Lett.\  {\bf 49}, 102 (1982).
  doi:10.1103/PhysRevLett.49.102


\bibitem{rajantie} O.~Gould and A.~Rajantie,
  Phys.\ Rev.\ D {\bf 96}, no. 7, 076002 (2017)
  doi:10.1103/PhysRevD.96.076002
  [arXiv:1704.04801 [hep-th]];
  arXiv:1705.07052 [hep-ph].



\bibitem{Schwinger} 
  J.~S.~Schwinger,
  Phys.\ Rev.\  {\bf 82}, 664 (1951).
  doi:10.1103/PhysRev.82.664



  

   \bibitem{susskind}  N.~Seiberg, L.~Susskind and N.~Toumbas,
  JHEP {\bf 0006}, 021 (2000)
  doi:10.1088/1126-6708/2000/06/021
  [hep-th/0005040].

\bibitem{mitsou} J.~R.~Ellis, K.~Farakos, N.~E.~Mavromatos, V.~A.~Mitsou and D.~V.~Nanopoulos,
  Astrophys.\ J.\  {\bf 535}, 139 (2000)
  doi:10.1086/308825
  [astro-ph/9907340].



\bibitem{mp} 
  N.~E.~Mavromatos and J.~Papavassiliou,
  Int.\ J.\ Mod.\ Phys.\ A {\bf 19}, 2355 (2004)
  doi:10.1142/S0217751X04018397
  [hep-th/0307028].

\bibitem{wu} T.-Y.~Wu and T.~Ohmura, \emph{Quantum Theory of Scattering} (Prentice Hall International Series in Physics, Prentice Hall Inc., Eaglewood Cliffs,  New Jersey (1962)). 

\bibitem{abram} M. Abramowitz and I.A. Stegun, eds. (1983), \emph{Handbook of Mathematical Functions with Formulas, Graphs, and Mathematical Tables}, 
Applied Mathematics Series. 55 (Washington D.C.; New York: United States Department of Commerce, National Bureau of Standards; Dover Publications. ISBN 978-0-486-61272-0. LCCN 64-60036). 

\bibitem{obc} J.~P.~M.~Pitelli, V.~S.~Barroso and M.~Richartz,
  Phys.\ Rev.\ D {\bf 96}, no. 10, 105021 (2017)
  doi:10.1103/PhysRevD.96.105021
  [arXiv:1711.03526 [gr-qc]].


 
\end{thebibliography}
\end{document}